\documentclass[final,3p,times,authoryear]{elsarticle}

\usepackage[utf8]{inputenc}
\usepackage[T1]{fontenc}
\usepackage{color}
\usepackage{algorithm2e}
\usepackage{float}
\usepackage{verbatim}
\usepackage{caption}
\usepackage{subcaption}
\usepackage{enumerate}
\usepackage{amsmath}
\usepackage{hyperref}
\usepackage{url}
\usepackage{cases}
\usepackage{amsthm}
\usepackage{multirow}
\usepackage{booktabs}

\newtheorem{theorem}{Theorem}

\newtheorem{lemma}{Lemma}
\newtheorem{corollary}{Corollary}
\newtheorem{definition}{Definition}
\newtheorem{result}{Result}
\newtheorem*{remark}{Remark}

\DeclareMathOperator{\rg}{rank}

\journal{arxiv}

\begin{document}

\begin{frontmatter}



\title{Nonlinear observability algorithms with known and unknown inputs:\\ analysis and implementation}


\author[1,2]{Nerea Martínez}
\author[1]{Alejandro F. Villaverde}
\ead{afvillaverde@iim.csic.es}
\address[1]{BioProcess Engineering Group, IIM-CSIC, Vigo 36208, Galicia, Spain}
\address[2]{Department of Applied Mathematics II, University of Vigo, Vigo 36310, Galicia, Spain}

\begin{abstract}
The observability of a dynamical system 
is affected by the presence of external inputs, either known (such as control actions) or unknown (disturbances). Inputs of unknown magnitude are especially detrimental for observability, and they also complicate its analysis. Hence the availability of computational tools capable of analysing the observability of nonlinear systems with unknown inputs has been limited until lately. Two symbolic algorithms based on differential geometry, ORC-DF and FISPO, have been recently proposed for this task, but their critical analysis and comparison is still lacking. Here we perform an analytical comparison of both algorithms and evaluate their performance on a set of problems, discussing their strengths and limitations. Additionally, we use these analyses to provide insights about certain aspects of the relationship between inputs and observability. We find that, while ORC-DF and FISPO follow a similar approach, they differ in key aspects that can have a substantial influence on their applicability and computational cost. The FISPO algorithm is more generally applicable, since it can analyse any nonlinear ODE model. The ORC-DF algorithm analyses models that are affine in the inputs, and if those models have known inputs it is sometimes more efficient. Thus, the optimal choice of a method depends on the characteristics of the problem under consideration. To facilitate the use of both algorithms we implement the ORC-DF algorithm in a new version of STRIKE-GOLDD, a MATLAB toolbox for structural identifiability and observability analysis. Since this software tool already had an implementation of the FISPO algorithm, the new release allows modellers and model users the convenience of choosing between different algorithms in a single tool, without changing the coding of their model.
\end{abstract}



\begin{keyword}
observability \sep identifiability \sep nonlinear systems \sep control theory \sep differential geometry \sep software


\end{keyword}

\end{frontmatter}


\section{Introduction}

Mathematical models of ordinary differential equations (ODEs) are used in all areas of science and technology for describing nonlinear systems. 
The ODEs define the derivatives of the state variables of the system with respect to time, $\dot x(t)$; the measurable quantities $y(t)$ are defined by the output function. The model equations (ODEs and output function) may contain unknown parameters, $\theta$, and external inputs that may be known ($u(t)$) or unknown ($w(t)$).
The structure of the model equations determines whether it is possible to estimate the model unknowns from the outputs. 
The theoretical possibility of inferring the states (respectively parameters) from the outputs is called observability (respectively structural identifiability) \citep{chatzis2015observability,villaverde2019observability}.
Since a parameter can be considered as a state variable with time derivative equal to zero, structural identifiability can be considered as a particular case of observability. The possibility of recovering the unknown inputs is called invertibility, reconstructibility, or input observability. For simplicity, in this manuscript we use the word \textit{observability} for all model unknowns, that is, to refer to the possibility of determining states, parameters, and/or inputs from the output.

The concept of observability arose in systems and control theory. It was initially defined for linear models and extended to the nonlinear case afterwards \citep{hermann1977nonlinear}. The concept of structural identifiability, on the other hand, was motivated by the analysis of biological models \citep{bellman1970structural}, due to the specific challenges that parameter identification poses in mathematical biology and other biosciences. Hence, many observability analysis methods developed in that context aimed at analysing structural identifiability and were named accordingly, even though they could be applied or adapted to the more general task of analysing observability. Examples of software tools include DAISY \citep{bellu2007daisy}, COMBOS \citep{meshkat2009algorithm}, EAR \citep{anguelova2012efficient}, STRIKE-GOLDD \citep{villaverde2016structural}, GenSSI \citep{ligon2018genssi}, and SIAN \citep{hong2019sian}.

The existence of external inputs affects the observability of a model and determines which methods can be applied for its analysis. A key distinction is between known and unknown inputs, where ``known'' is interpreted as ``quantified''; thus, we are aware of the existence of an unknown input but not of its magnitude.
A known input that can be manipulated is also called a control input, or simply a control. An unknown input can be considered as an unmeasured disturbance or as a time-varying parameter. 
Some techniques are applicable specifically to uncontrolled systems \citep{evans2002identifiability}, while others allow for the existence of known inputs. Few methods are capable of handling both known and unknown inputs. Some exceptions \citep{martinelli2015extension,martinelli2018nonlinear} are not applicable to systems in which the outputs are a direct function of the inputs, and do not analyse the observability of the unknown input itself.

Two differential geometry algorithms 
called ORC-DF -- Observability Rank Condition with Direct Feedthrough \citep{maes2019observability} -- and FISPO -- Full Input, State, and Parameter Observability \citep{villaverde2019full} -- have been recently presented. Both methods are capable of determining the observability of states, parameters, and inputs of nonlinear ODE models. ORC-DF is applicable to affine-in-the-inputs systems, while FISPO does not have this requirement. The FISPO algorithm is implemented in the STRIKE-GOLDD toolbox \citep{villaverde2016structural}.

In the present paper we perform a critical examination of the ORC-DF and FISPO algorithms. First we provide the necessary background on observability analysis and differential geometry in Section \ref{sec:methods}. Then we perform a theoretical analysis of the two methods in Section \ref{sec:theory}, showing that they are equivalent for a certain class of problems and describing how they differ for other classes. 
Realizing the convenience of having both algorithms available in the same software environment, we provide their implementations in a new version of the MATLAB toolbox STRIKE-GOLDD, which is described in Section \ref{sec:implementation}. The new release includes an implementation of ORC-DF, as well as a seamless integration with the already existing FISPO.  
Furthermore, it enables the automatic analysis of multi-experiment observability.
Since ORC-DF and FISPO are symbolic algorithms that can be computationally expensive, in Section \ref{sec:results} we evaluate their performance by applying them to a number of modelling problems of different domains, from mechanical engineering to biology, and report their applicability and computational cost. The analysis of the selected case studies is also helpful for obtaining detailed insights about the inner working of the algorithms. Finally, we conclude with a discussion of the results in in Section \ref{sec:conclusion}.

\section{Materials and methods}\label{sec:methods}

\subsection{Notation and model classes}

We are interested in the observability of nonlinear systems of the following form: 
\begin{numcases}{\Sigma=}\label{din}
	\dot{x}(t)=f\left(x(t),\theta,u(t),w(t)\right)\\\label{out}
	y(t)=h\left(x(t),\theta,u(t),w(t)\right)
\end{numcases}
defined for all $t\in I,$ $I\subset \left[0,+\infty\right)$ an arbitrary time interval, where $f$ and $h$ are nonlinear and analytical (infinitely differentiable) functions of the states $x(t)\in\mathbb{R}^{n_x},$ known and unknown inputs, $u(t)\in\mathbb{R}^{n_u}$ and $w(t)\in\mathbb{R}^{n_w}$ respectively, and unmeasured parameters $\theta\in\mathbb{R}^{n_p}.$ The output $y(t)\in\mathbb{R}^m$ consists of measurement functions of model variables.

As a special case of 
(\ref{din}--\ref{out}) we also study systems affine in the inputs, which are of the following form:
\begin{numcases}{\Sigma_A=}\label{dinaffine}
\dot{x}(t)=f_0\left(x(t),\theta\right)+\sum_{i=1}^{n_u}f_{u_i}\left(x(t),\theta\right)u_i(t)+\sum_{i=1}^{n_w}f_{w_i}\left(x(t),\theta\right)w_i(t)\\\label{outaffine}
y(t)=h_0\left(x(t),\theta\right)+\sum_{i=1}^{n_u}h_{u_i}\left(x(t),\theta\right)u_i(t)+\sum_{i=1}^{n_w}h_{w_i}\left(x(t),\theta\right)w_i(t)
\end{numcases}
where $f_0,\;f_{u_i},\;f_{w_j},\;h_0,\;h_{u_i},\;h_{w_j}$, for $1\leq i\leq n_u, 1\leq j\leq n_w$, are analytical functions -- possibly nonlinear -- and:
\begin{align}
\label{dynafin}
&f=f_{xw}+\sum_{i=1}^{n_u}f_{u_i}u_i,\quad h=h_{xw}+\sum_{i=1}^{n_u}h_{u_i}u_i,
\end{align}
where, following \cite{maes2019observability}:
\begin{align}\label{dynafin2}
&f_{xw}=f_0+\sum_{i=1}^{n_w}f_{w_i}w_i,\quad h_{xw}=h_{0}+\sum_{i=1}^{n_w}h_{w_i}w_i.\end{align}

In what follows, a vector $v\in\mathbb{R}^n$ is assumed to be a one column matrix and $v^T$ its transpose. The Jacobian matrix of a function $\phi=\left(\phi_1,\dots,\phi_s\right)$ 
with respect to a vector field $v=\left(v_1,\dots,v_n\right),$ will be denoted as:
\begin{align*}
\frac{\partial \phi}{\partial v}=\left[\frac{\partial \phi_i}{\partial v_j}\right]_{ij}\quad 1\leq i\leq s,\;1\leq j\leq n.
\end{align*}
\subsection{Background
}

\subsubsection{Structural identifiability, observability, and differential geometry}

A nonlinear system $\Sigma$ is structurally observable if it is possible to distinguish between its state trajectories from the data provided by its output, and structurally reconstructible (or invertible) if its disturbances can be tracked from the aforementioned measurements. Similarly, $\Sigma$ is structurally identifiable if it is possible to infer the values of its unknown parameters from the output.
In practice it is often not necessary distinguish between every pair of unmeasured states in the phase mapping of $\Sigma$ -- a property called structural global observability -- and it is sufficient to distinguish neighbouring states -- a property called local ``weak'' observability in some texts \citep{hermann1977nonlinear}. In this work we will not make this distinction, and local ``weak'' structural observability will be simply called observability. Likewise, we will refer to structural local identifiability and structural local invertibility simply as identifiability and invertibility.

Structural identifiability can be studied as particular case of observability. Since any unknown parameter $\theta_i$ of $\Sigma$ can be considered as a constant state, that is, $\dot\theta_i=0$ holds for $1\leq i\leq n_p,$ it is possible to augment the state vector as: \begin{align}\label{x_par}
\tilde{x}=\begin{pmatrix}x&\theta\end{pmatrix}^T\end{align} 
which consists of $n_{\tilde{x}}=n_x+n_\theta$ components and follows the augmented dynamics:
\begin{align}\label{sys_aug}
\dot{ \tilde{x}}(t)=
\begin{pmatrix}
\dot x(t)&
\dot {\theta}
\end{pmatrix}^T=
\begin{pmatrix}
f\left(x(t),\theta,u(t),w(t)\right)&
0_{1\times n_p}
\end{pmatrix}^T=\tilde{f}\left(\tilde{x}(t),u(t),w(t)\right)
\end{align}
Thus, the identifiability and observability of $\Sigma$ can be studied as the observability of the augmented system with states \eqref{x_par}, dynamics \eqref{sys_aug}, and the same output as $\Sigma.$ 

The algorithms analysed in this work adopt a differential geometry approach, which uses the concept of Lie derivative to bring out algebraic conditions that establish observability. 
Let us consider first the case in which $\Sigma$ is not dependent on unknown inputs, that is:
\begin{align*}
\Sigma'\begin{cases}
&\dot{x}(t)=f\left(x(t),\theta,u(t)\right)\\
&y(t)=h\left(x(t),\theta,u(t)\right)
\end{cases}
\end{align*}

\begin{definition}[Lie derivative \citep{vidyasagar1993nonlinear}]\label{lie_der}
Consider the system $\Sigma'$ with augmented state vector \eqref{x_par} and augmented dynamics \eqref{sys_aug}, and assume that the inputs $u(t)$ are constant. The Lie derivative of the output function $h$ along the tangent vector field $\tilde f=\tilde{f}\left(\cdot,u\right)$ is:
	\begin{align*}
	&L_{\tilde{f}}h\left(\tilde{x}(t),u(t)\right)=\frac{\partial h}{\partial \tilde{x}}\left(\tilde{x}(t),u(t)\right) \tilde{f}\left(\tilde{x}(t),u(t)\right),
	\end{align*}
	and, setting $L^0_{\tilde{f}}h=h,$ the $i-$order Lie derivative can be recursively computed as:
	\begin{align*}
	&L_{\tilde{f}}^ih\left(\tilde{x}(t),u(t)\right)=L_{\tilde{f}}\left(L_{\tilde{f}}^{i-1}h\left(\tilde{x}(t),u(t)\right)\right),\quad i\geq 1.
	\end{align*}
\end{definition}
The above definition can be extended to the case of analytical inputs as follows:
\begin{definition}[Extended Lie derivative \citep{anguelova2012efficient}]\label{lie_ext}
	Consider the system $\Sigma'$ with augmented state vector \eqref{x_par}, augmented dynamics $\eqref{sys_aug}$, and assume that the inputs $u(t)$ are analytical functions. The extended Lie derivative of the output function $h$ by the tangent vector field $\tilde{f}=\tilde f\left(\cdot,u\right)$ is:
	\begin{align*}
	&L_{\tilde{f}}^eh\left(\tilde{x}(t),u(t)\right)=\frac{\partial h}{\partial \tilde{x}}\left(\tilde{x}(t),u(t)\right) \tilde{f}\left(\tilde{x}(t),u(t)\right)+\frac{\partial h}{\partial u}\left(\tilde{x}(t),u(t)\right) \dot{u}(t)
	\end{align*}
	and, setting $L^{e,0}_{\tilde{f}}h=h,$ the $i-$order extended Lie derivative can be recursively computed as:
	\begin{align*}
	&L^{e,i}_{\tilde{f}}h\left(\tilde{x}(t),u(t)\right)=\frac{\partial L^{e,i-1}_{\tilde{f}}h}{\partial \tilde{x}}\left(\tilde{x}(t),u(t)\right) \tilde{f}\left(\tilde{x}(t),u(t)\right)+\sum_{j=0}^{i-1}\frac{\partial L^{e,i-1}_{\tilde{f}}h}{\partial u^{\left.j\right)}}\left(\tilde{x}(t),u(t)\right) u^{\left.j+1\right)}(t),\quad i\geq 1.
	\end{align*}
\end{definition}
\begin{remark}
Note that, in the case of constant inputs, the Lie derivative introduced in Definition \eqref{lie_der} verifies
\begin{align*}
    &y(t)=h\left(\tilde x(t),u\right)=L^0_{\tilde f}h\left(\tilde x(t),u\right)\\
    &y'(t)=\frac{d}{dt}h\left(\tilde x(t),u\right)=\frac{\partial h}{\partial \tilde x}\left(\tilde x(t),u\right)\dot{\tilde x}(t)=\frac{\partial h}{\partial \tilde x}\left(\tilde x(t),u\right)\tilde f\left(\tilde{x}(t),u(t)\right)=L_{\tilde f}h\left(\tilde x(t),u\right)\\
    &y''(t)=\frac{d}{dt}y'(t)=\frac{d}{dt}L_{\tilde f}h\left(\tilde x(t),u\right)=\frac{\partial L_{\tilde f}h}{\partial \tilde x }\left(\tilde{x}(t),u(t)\right)\dot{\tilde x}(t)=\frac{\partial L_{\tilde f}h}{\partial \tilde x }\left(\tilde{x}(t),u(t)\right)\tilde f\left({\tilde x}(t),u\right)=L^2_{\tilde f}h\left({\tilde x}(t),u\right)\\
    &\begin{matrix}
    &\vdots\\
    \end{matrix}\\
    &y^{\left.i\right)}(t)=\frac{d}{dt}L_{\tilde f}^{i-1}h\left(\tilde x(t),u\right)=L_{\tilde f}\left(L_{\tilde f}^{i-1}h\left(\tilde x(t),u\right)\right)=L_{\tilde f}^ih\left(\tilde x(t),u\right),\quad i\geq 0.
\end{align*}
by using repeatedly the chain rule. Likewise, the extended Lie derivative of Definition \eqref{lie_ext} verifies:
\begin{align*}
    &y^{\left. i\right)}(t)=L^{e,i}_{\tilde f}h\left(\tilde x(t),u(t)\right),\quad i\geq 0.
\end{align*}
\end{remark}
Given a nonlinear system $\Sigma'$ with augmented state \eqref{x_par} and analytical inputs, it is possible to use the extended Lie derivatives of the output to build the following $mn_{\tilde{x}}\times n_{\tilde{x}}$ matrix:
\begin{align}\label{matr_obs}
&\mathcal{O}_I\left(\tilde{x},u\right)=\frac{\partial}{\partial \tilde{x}}
\begin{pmatrix}
L^0_{\tilde{f}}h\left(\tilde{x},u\right)^T&L_{\tilde{f}}h\left(\tilde{x},u\right)^T&L^2_{\tilde{f}}h\left(\tilde{x},u\right)^T&\dots& L^{n_{\tilde{x}}-1}_{\tilde{f}}h\left(\tilde{x},u\right)^T
\end{pmatrix}^T,
\end{align}
which is the observability-identifiability matrix of $\Sigma'.$ By calculating the rank of the above matrix, it is possible to establish the observability and identifiability of $\Sigma'$ using the following condition.
\begin{theorem}[Observability-identifiability condition, OIC \citep{anguelova2012efficient}]\label{OIC}
	If the identifiability-observability matrix of a model $\Sigma'$ satisfies $\rg\left(\mathcal{O}_I\left(\tilde{x}_0,u\right)\right)=n_{\tilde{x}},$ with $\tilde{x}_0$ being a (possibly generic) point in the augmented state space \eqref{sys_aug} of $\Sigma'$, then the system is structurally locally observable and structurally locally identifiable.
\end{theorem}
\begin{remark}
 The rank of $\mathcal{O}_I$ is constant except for a zero-measurement subset in the augmented state space \eqref{sys_aug} of $\Sigma'$ where the rank is smaller, as a consequence of the system being analytical \citep{isidori1995nonlinear}. Thus, to verify the condition of the Theorem \ref{OIC} it is sufficient to calculate the rank of $\mathcal{O}_I$ at any non-singular point of the phase space.
\end{remark}

\subsubsection{FISPO}

The effect of unknown inputs $w$ can be taken into account by further augmenting $\Sigma$, including $w$ as unmeasured states. Thus, for a non-negative integer $l$ we have the $l-$augmented states vector:
\begin{align}\label{x_aug}
&x^l=\begin{pmatrix}
x^T&
\theta^T&
w^T&
\dots&
w^{\left.l\right)^T}\end{pmatrix}^T,
\end{align}
which follows the $l-$augmented dynamics:
\begin{align*}
&\dot x^l(t)=f^l\left(x^l(t),u(t),w^{\left.l+1\right)}(t)\right)=\begin{pmatrix}
f\left(x^0(t),u(t)\right)^T&
0_{1\times n_p}&
w(t)^T&
\dots&
w^{\left.l+1\right)}(t)^T
\end{pmatrix}^T,
\end{align*}
leading to the $l-$augmented system:
\begin{align}\label{aug_sys}
\Sigma^l
&\begin{cases}
&\dot x^l(t)=f^l(x^l(t),u(t),w^{\left.l+1\right)}(t))\\
&y(t)=h(x^0(t),u(t))
\end{cases}
\end{align}

An analogous extension for affine systems $\Sigma_A$ exists. Using the notation given in \eqref{dynafin}--\eqref{dynafin2}--\eqref{x_par}, the $l-$augmented system described above takes the form \citep{maes2019observability}:
\begin{align*} 
&\Sigma_A^l\begin{cases}
&\dot{x}^l(t)=f_{xw}^l(x^l(t),w^{\left.l+1\right)}(t))+\sum_{i=1}^{n_u}f^l_{u_i}\left(\tilde{x}(t)\right)u_i(t)\\
&y(t)=h_{xw}(x^0(t))+\sum_{i=1}^{n_u}h_{u_i}\left(\tilde{x}(t)\right)u_i(t)
\end{cases}
\end{align*}
where the $l-$augmented dynamics is decomposed as follows:
\begin{align}\label{input_dyn_am}
&f_{xw}^l\left(x^l(t),w^{\left.l+1\right)}(t)\right)=\begin{pmatrix}
f_{xw}\left(x^0(t)\right)^T&
0_{1\times n_p}&
\dot{w}(t)^T&
\dots&
w^{\left.l+1\right)}(t)^T
\end{pmatrix}^T\\\label{input_dyn_am2}
&f^l_{u_i}\left(\tilde{x}(t)\right)=\begin{pmatrix}
f_{u_i}\left(\tilde{x}(t)\right)&
0_{1\times n_p}&
0_{1\times(l+1)n_w}\end{pmatrix}^T,\quad 1\leq i\leq n_u.
\end{align}

We note that, in order to build $l-$augmented systems $\Sigma^l$ and $\Sigma_A^l,$ it must be possible to calculate the $l+1-$time derivative of disturbances $w(t)$ and, therefore, they will be considered as analytical functions from now on. We also note that the $l-$augmented form of $\Sigma$ and $\Sigma_A$ is equivalent to the original system, which consists of $n^l=n_x+n_\theta+\left(l+1\right)n_w$ states, $n_u$ controls, $n_w$ disturbances (the $l+1-$order time derivatives of $\left.w\right)$ and $m$ outputs, that have not changed due to state augmentation \citep{martinelli2015extension}.

As an additional hypothesis we assume that a non-negative integer $s$ exists (possibly $\left.s=+\infty\right)$ such that $w^{\left.s\right)}(t)\neq 0$ and $w^{\left.i\right)}(t)=0$ for all $i>s.$ In principle, this assumption introduces a restriction on the type of allowed inputs, and it is equivalent to assuming that the disturbances are polynomial functions of time. Nevertheless, in practice, the method may still provide relevant information about the general case, as is discussed in \citep{villaverde2019full}.

In what follows, if a vector function $\phi$ depends on variables $x^l$ we denote:
\begin{align*}
&d^l\phi(x^l)=\frac{\partial\phi}{\partial x^l}(x^l)\\
&L_{f^l}\phi(x^{l+1})=d^l\phi(x^l) f^l(x^{l+1})
\end{align*}
and, if $\phi=\phi\left(\cdot,u\right)$ (where controls $u(t)$ are considered to be analytical) then:
\begin{align*}
&L^e_{f^l}\phi(x^{l+1},u)=d^l\phi(x^l,u) f^l(x^{l+1},u)+\frac{\partial \phi}{\partial u}(x^l,u)\dot{u}
\end{align*}
\begin{definition}[Full Input-State-Parameter
	Observability, FISPO \citep{villaverde2019full}]
	Consider the system $\Sigma$ and the augmented states vector $z(t)=\left(
	x(t),\theta,w(t)
	\right).$ We say that $\Sigma$ has the FISPO property if, for every $t_0\in I$ and $1\leq i\leq n^0,$ $z_i(t_0)$ can be determined from the output $y(t)$ and the known inputs $u(t)=\left(u_1(t),\dots,u_{n_u}(t)\right)$ in a finite time interval $\left[t_0,t_f\right]\subset I.$ Thus, a system $\Sigma$ is FISPO if, for every $z(t_0)$ and for almost any vector $z^\ast(t_0),$ there is a neighbourhood $\mathcal{N}\left(z^\ast\left(t_0\right)\right)$ such that, for all $\hat{z}(t_0)\in\mathcal{N}\left(z^\ast\left(t_0\right)\right),$ the following condition holds:
	\begin{align*}
	&y\left(t,\hat{z}(t_0)\right)=y\left(t,z^\ast\left(t_0\right)\right)\Rightarrow \hat{z}_i\left(t_0\right)=z_i^\ast\left(t_0\right),\quad 1\leq i \leq n^0.
	\end{align*}
\end{definition}

\begin{remark}
The original definition of the term FISPO reproduced above refers to a model property. Here we also use it to refer to the algorithm presented for its evaluation by  \cite{villaverde2019full}. 
\end{remark}

Using the system augmentation \eqref{aug_sys} and taking the unique $l=s$ such that $w^{\left.s\right)}(t)\neq 0$ and $w^{\left. i\right)}(t)=0$ for all $i>s,$ it is possible to build the following matrix,
\begin{align}\label{gen-matrix}
&\mathcal{O}^g_I\left(x^s,u\right)=d^s\begin{pmatrix}
L^0_{f^s}h\left(x^s,u\right)^T&L_{f^s}h\left(x^s,u\right)^T&L^2_{f^s}h\left(x^s,u\right)^T&\dots&L^{n^s-1}_{f^s}h\left(x^s,u\right)^T
\end{pmatrix}^T,
\end{align}
which is the generalized observability matrix of $\Sigma.$ Note that \eqref{gen-matrix} coincides with the observability matrix \eqref{matr_obs} of $\Sigma^s$ without disturbances. Thus, the rank of $\mathcal{O}^g_I$ provides a condition for assessing the observability of $\Sigma$ as follows:
\begin{theorem}[FISPO Condition \citep{villaverde2019full}]
A nonlinear system $\Sigma$ given by (\ref{din}-\ref{out}) with analytic inputs is FISPO if, for $x^s_0$ being a (possibly generic) point in the state space of the $s-$augmented system $\Sigma^s,$ the generalized observability matrix (\ref{gen-matrix}) verifies $\rg\left(\mathcal{O}^g_I(x^s_0,u)\right)=n^s.$
\end{theorem}
\begin{remark}
For $1\leq i\leq n^s,$ the observability of the $i-$th state of $x^s$ can also be studied using the matrix $\eqref{gen-matrix}.$ Thus, if $\mathcal{O}^{g,i}_I(x_0^s,u)$ denotes the matrix obtained from $\mathcal{O}_I^g(x_0^s,u)$ after removing its $i-$th column, state $x_i$ is observable if $\rg\left(\mathcal{O}_I^{g,i}(x_0^s,u)\right)<\rg\left(\mathcal{O}_I^g(x_0^s,u)\right)$ for almost any $x_0^s$ in the phase space of $\Sigma^s.$  
\end{remark}

\subsubsection{ORC-DF}

The observability of affine systems $\Sigma_A$ with bounded controls $u(t)$ can also be analysed by building a different observability matrix, as explained below. For a full description of the procedure, see \citep{maes2019observability}.
\begin{definition}[Observability Rank Criterion for systems with Direct Feedthrough, ORC-DF]
	A system $\Sigma_A$ is classified as $k-$row observable if almost any initial state $x^k(t_0),$ $t_0\in I,$ in the state space of the $k-$augmented system $\Sigma_A^k$ can be separated locally from its neighbours based on the output at $k+1$ consecutive times $t_0,$ $t_1,$ $\dots,$ $t_k.$ If there exists $k\geq 1$ such that $\Sigma_A$ is $k-$row observable, it is said that $\Sigma_A$ satisfies the ORC-DF.
\end{definition}
\begin{lemma}\label{orc-df}
	Consider the system $\Sigma_A$ and the vector field $\Omega_k,$ which is recursively defined by
	\begin{align*}
	&\Omega_0=\begin{pmatrix}
	h_{xw}^T&
	h_{u_1}^T&
	\dots&
	h_{u_{n_u}}^T
	\end{pmatrix}^T,\quad 
	\Delta\Omega_0=\Omega_0,\\
	& \Omega_{k+1}=\begin{pmatrix}
	\Omega_k^T&
	\Delta \Omega_{k+1}^T
	\end{pmatrix}^T,\quad\Delta\Omega_{k+1}=\begin{pmatrix}
	L_{f^k_{xw}}\left(\Delta\Omega_k\right)^T&
	L_{f^k_{u_1}}\left(\Delta\Omega_k\right)^T&
	\dots&
	L_{f^k_{u_{n_u}}}\left(\Delta\Omega_k\right)^T
	\end{pmatrix}^T,\quad k\geq 0,
	\end{align*}
	then $\Sigma_A$ is $k-$row observable if $\rg\left(d^k\Omega_k\left(x^k_0\right)\right)=n^k$ for almost any $x_0^k$ in the phase space of $\Sigma_A^k.$
\end{lemma}
\begin{lemma} 
	If $\Sigma_A$ satisfies the ORC-DF, then $\Sigma_A$ is observable in the presence of unmeasured inputs.
\end{lemma}
\begin{corollary}
	Let $d^k\Omega_k^i$ denote the matrix that is obtained after removing the $i-$th column from $d^k\Omega_k.$ The $i-$th state of $x^k$ is $k-$row observable if and only if $\rg\left(d^k\Omega_k^i\left(x_0^k\right)\right)<\rg\left(d^k\Omega_k\left(x_0^k\right)\right)$ for almost any $x_0^k$ in the phase space of $\Sigma^k.$
\end{corollary}

\section{Theory: analysis of the FISPO and ORC-DF algorithms}\label{sec:theory}

In this section we discuss the similarities and differences between FISPO and ORC-DF, whose pseudo-code is provided in Algorithms \ref{fispoalg}--\ref{orc-alg}.

\subsection{Preliminary remarks}

We begin by recalling three facts that are relevant for the analysis: (i) the FISPO algorithm does not always require building the full matrix, (ii) both ORC-DF and FISPO can be inconclusive for certain models, (iii) ORC-DF and FISPO can handle different types of inputs.

\begin{remark}[The FISPO algorithm does not always require building the full matrix]
In each iteration the FISPO algorithm builds the matrix $\mathcal{O}_I^k(x^k,u),$ composed by extended Lie derivatives of output up to order $k,$ and then calculates its rank and partial ranks, instead of directly building the full matrix \eqref{gen-matrix}. The algorithm is programmed in this way because the matrix $\mathcal{O}_I^k(x^k,u)$ can reach full rank for some $k\leq n_s-1$, and if the number of states increases indefinitely, \eqref{gen-matrix} can never be built in practice. In addition, the above procedure may classify some states as observable before obtaining the full matrix \eqref{gen-matrix}, since any observable state in the $k-$ augmented system $\Sigma^k$ remains observable in $\Sigma^l,$ for $l\geq k$ \citep{martinelli2015extension}. Moreover, if the system does not have unknown inputs it is possible to classify it as unobservable or unidentifiable using fewer than $n_{\tilde{x}}-1$ Lie derivatives \citep{anguelova2004nonlinear}.
\end{remark}

\begin{algorithm}[H]
	\SetAlgoLined
	\KwResult{Observable and unobservable states, parameters and disturbances.}
	$k=0,\quad x^0=\begin{pmatrix}
	x^T&\theta^T&w^T
	\end{pmatrix}^T,\quad n^0=n_x+n_\theta+n_w,\quad \Lambda^0=\left\lbrace 1,2,\dots,n^0\right\rbrace$\;
	$f^0=\begin{pmatrix}
	f^T&0_{1\times n_\theta}&\dot{w}^T
	\end{pmatrix}^T$\;
	$\mathcal{O}_I^0=d^0h$\;
	\While{$\rg\left(\mathcal{O}_I^k\right)<n^k$}{
		$k=k+1$\;
		$x^k=\begin{pmatrix}x^{k-1^T}&w^{\left.k\right)^T}\end{pmatrix}^T,\quad n^k=n^{k-1}+n_w,\quad \Lambda^k=\Lambda^{k-1}\cup\left\lbrace n^{k-1}+1,n^{k-1}+2,\dots,n^k\right\rbrace$\;
		$\mathcal{O}_I^k=\begin{pmatrix}
		\mathcal{O}_I^{k-1}&0\\
		&d^kL_{f^{k-1}}^{e,k}h
		\end{pmatrix}$\;
			\For {$i\in \Lambda^k$}{
			$\mathcal{O}_I^{k,i}=\mathcal{O}_I^k-\left\lbrace\text{i column}\right\rbrace$\;
			\If{
				$\rg\left(\mathcal{O}_I^{k,i}\right)<\rg\left(\mathcal{O}_I^k\right)$}{
				$x_i$ is observable\;
				$\Lambda^k=\Lambda^k-\left\lbrace i\right\rbrace$\;}
		}
		$f^k=\begin{pmatrix}
		f^{k-1^T}&w^{\left.k+1\right)^T}
		\end{pmatrix}^T$\;
	}
	\caption{The FISPO algorithm \citep{villaverde2019full}.}
	\label{fispoalg}
\end{algorithm}

\vspace{1cm} 

\begin{algorithm}[H]
	\SetAlgoLined
	
	\KwResult{Observable and unobservable states, parameters and disturbances.}
	$k=0,\quad x^0=\begin{pmatrix} x^T&\theta^T&w^T\end{pmatrix}^T,\quad n^0=n_x+n_\theta+n_w,\quad \Lambda^0=\left\lbrace 1,2,\dots,n^0\right\rbrace$\;
	$f_{xw}^0=\begin{pmatrix}f_{xw}^T&0_{1\times n_\theta}&\dot{w}^T\end{pmatrix}^T,\quad f_{u_i}^0=\begin{pmatrix}
	f_{u_i}^T&0_{1\times n_\theta}&0_{1\times n_w}
	\end{pmatrix}^T,\quad 1\leq i\leq n_u$\;
	$\Delta\Omega_0=\begin{pmatrix}
	h_{xw}^T&h_{u_1}^T&\dots&h_{u_{n_u}}^T\end{pmatrix}^T,\quad \Omega_0=\Delta\Omega_0$\;
	
	\While{$\rg\left(d^k\Omega_k\right)<n^k$}{
		$k=k+1$\;
		$\Delta\Omega_k= \begin{pmatrix}
		L_{f^{k-1}_{xw}}\left(\Omega_{k-1}\right)^T&
		L_{f^{k-1}_{u_1}}\left(\Omega_{k-1}\right)^T&
		\dots&
		L_{f^{k-1}_{u_{n_u}}}\left(\Omega_{k-1}\right)^T
		\end{pmatrix}^T,\quad \Omega_k=\begin{pmatrix}
		\Omega_{k-1}^T&\Delta \Omega_k^T
		\end{pmatrix}^T$\;
		$x^k=\begin{pmatrix}x^{k-1^T}&w^{\left.k\right)^T}\end{pmatrix}^T,\quad n^k=n^{k-1}+n_w,\quad \Lambda^k=\Lambda^{k-1}\cup\left\lbrace n^{k-1}+1,n^{k-1}+2,\dots,n^k\right\rbrace$\;
		$d^k\Omega_k=\begin{pmatrix}
		d^{k-1}\Omega_{k-1}&0\\
		&d^k\Delta\Omega_k\end{pmatrix}$\;
		\For {$i\in \Lambda^k$}{
			$d^k\Omega_k^i=d^k\Omega_k-\left\lbrace i \text{ column}\right\rbrace$\;
			\If{
				$\rg\left(d^k\Omega_k^i\right)<\rg\left(d^k\Omega_k\right)$}{
				$x_i$ is $k-$row observable\;
				$\Lambda^k=\Lambda^k-\left\lbrace i\right\rbrace$\;}
		}
		$f_{xw}^k=\begin{pmatrix}f_{xw}^{k-1^T}&w^{\left.k+1\right)^T}\end{pmatrix}^T,\quad f_{u_i}^k=\begin{pmatrix}
		f_{u_i}^{k-1^T}&0_{1\times n_w}
		\end{pmatrix}^T,\quad 1\leq i\leq n_u$\;
	}
	\label{orc-alg}
	\caption{The ORC-DF algorithm \citep{maes2019observability}.}
\end{algorithm}

\clearpage

\begin{remark}[Both ORC-DF and FISPO can be inconclusive for certain models]
If the model under study has unknown inputs and their time derivatives $w^{\left.j\right)}(t)$ do not vanish for any non-negative integer $j<+\infty,$ both FISPO and ORC-DF algorithms can be inconclusive. This happens when the rank of the observability matrices grows at each iteration without reaching a value equal to the number of states (which also increases with each iteration). Therefore, a computational implementation of both algorithms should include shutdown conditions based on computation time or number of iterations.
\end{remark}

\begin{remark}[ORC-DF and FISPO can handle different types of inputs]
Both ORC-DF and FISPO construct an observation space  generated by Lie derivatives of the output; its dimension determines observability. FISPO builds an observation space spanned by extended Lie derivatives \eqref{lie_ext} considering analytical inputs, while ORC-DF assumes piecewise constant inputs and exploits certain properties specific to affine systems in order to build a different observation space. If an affine system is classified as observable by ORC-DF or FISPO, it is observable when a generic measurable input is considered \citep{anguelova2004nonlinear,maes2019observability}.
\end{remark}

In the next subsections we present the main novel insights of our theoretical analysis of the algorithms.

\subsection{For systems without known inputs, ORC-DF and FISPO reduce to the same algorithm}\label{fispo=orc}

Here we prove by induction that, if no inputs $u$ are involved in $\Sigma_A,$ the FISPO algorithm reduces to ORC-DF. Before presenting the result, we remark that, in the case $n_u=0,$ the extended Lie derivative reduces to:
\begin{align*} &L^e_f\left(\cdot\right)=L_f\left(\cdot\right)\end{align*}
and, denoting the composition of functions with $\circ$, the $k+1-$order Lie derivative verifies:
\begin{align}\label{rel_lie}
&L_{f^{k}}^{k+1}(h)=L_{f^{k}}\circ\overbrace{\cdots}^{k+1}\circ\; L_{f^{k}}\left(h\right)=L_{f^{k}}\circ L_{f^{k-1}}\circ\cdots\circ L_{f^0}\left(h\right)=L_{f^{k}}\left(L^{k}_{f^{k-1}}h\right)\quad k\geq 1,
\end{align}
since $L^j_{f^{k}}h$ depends only on time derivatives $w^{\left.i\right)}(t)$ for $1\leq i\leq j\leq k+1.$
\begin{result}
	If the system $\Sigma_A$ is independent of any known input, then $d^k\Omega_k=\mathcal{O}_I^k$ for all $k\geq 0.$
\end{result}
\begin{proof}
	Setting $n_u=0$ in \eqref{dynafin}, the dynamics and output of $\Sigma_A$ are given by:
	\begin{align*}
	&f\left(x(t),\theta,w(t)\right)=f_{xw}\left(x(t),\theta,w(t)\right)\\
	&h\left(x(t),\theta,w(t)\right)=h_{xw}\left(x(t),\theta,w(t)\right)
	\end{align*}
	Let $k=0.$ By the recursion given in Lemma \eqref{orc-df}, it is verified that:
	\begin{align}\label{delta0}
	&\Delta\Omega_0=h_{xw}=h,
	\end{align}
	so the induction hypothesis holds for $k=0\colon$
	\begin{align*}
	&d^0\Omega_0=d^0\Delta\Omega_0=d^0
	h=\mathcal{O}^0_I.
	\end{align*}
	
	Consider now any non-negative integer $k\geq 0$ and suppose that the induction hypothesis holds for $0\leq j\leq k,$ then:
	\begin{align*}
	&d^{k+1}\Omega_{k+1}=d^{k+1}\begin{pmatrix}
	\Omega_{k}^T&\Delta\Omega_{k+1}^T
	\end{pmatrix}^T=\begin{pmatrix}
	d^{k}\Omega_{k}&0\\
	&d^{k+1}\Delta\Omega_{k+1}
	\end{pmatrix}=\begin{pmatrix}
	\mathcal{O}_I^{k}&0\\
	&d^{k+1}\Delta\Omega_{k+1}
	\end{pmatrix}
	\end{align*}
	and the result is proven if for every $k\geq 0$ it holds that:
	\begin{align}\label{kmasuno}
	&\Delta\Omega_{k+1}=L^{e,k+1}_{f^{k}}h=L^{k+1}_{f^{k}}h,
	\end{align}
	
	The above equality is fulfilled for $k\geq 0.$ 
	Indeed, for $k=0,$ using \eqref{delta0} we have:
	\begin{align*}
	&\Delta\Omega_1=L_{f^0_{xw}}\left(\Delta\Omega_0\right)=L_{f_{xw}^0}h=L_{f^0}h
	\end{align*}
	and, if $k\geq 0$ and the condition \eqref{kmasuno} holds for $0\leq j\leq  k-1,$ then:
	\begin{align*}
	\Delta\Omega_{k+1}=L_{f_{xw}^{k}}\left(\Delta\Omega_{k}\right)=L_{f^{k}}\left(\Delta\Omega_{k}\right)=L_{f^{k}}\left(L^{k}_{f^{k-1}}h\right)=L^{k+1}_{f^k}h
	\end{align*}
	where in the last equality we have applied \eqref{rel_lie}. Thus, condition \eqref{kmasuno} holds for $k\geq 0.$
\end{proof}

\subsection{For systems with known inputs, ORC-DF and FISPO lead to different observability matrices}\label{lab_section}

Excluding the case $n_u=0,$ an important difference between the observability matrices built by algorithms ORC-DF and FISPO is the number of Lie derivatives (rows) they include in each iteration. Indeed, for $k\geq 0,$ 
\begin{align*}
&\mathcal{O}_I^k(x^k,u)\in m\left(k+1\right)\times n^k,\quad d^k\Omega^k(x^k)\in \sum_{i=0}^{k}m\left(1+n_u\right)^{i+1}\times n^k,
\end{align*}
so the observability matrix constructed by FISPO grows in $m$ rows in each iteration, while the matrix constructed by ORC-DF includes $m\left(1+n_u\right)^{k+1}$ new rows in the $k-$th stage. This fact can be an advantage for ORC-DF, as it makes it possible to reach full rank more rapidly, i.e. with lower order Lie derivatives. However, it may also be a disadvantage if this growth makes the problem dimension increase rapidly while adding little new information. Hence the faster growth may be beneficial or not depending on the form of the mathematical expressions of the dynamics and output functions in which the known inputs are present.
For example, suppose that there exists an integer $1\leq i\leq n_u$ and $n_u-1$ real numbers $\lambda_j$ not simultaneously zero, such that:
\begin{align*}
& f_{u_i}=\sum_{j\neq i=1}^{n_u}\lambda_j f_{u_j}
\end{align*}
which, using \eqref{input_dyn_am}, implies:
\begin{align*}
& f^k_{u_i}=\sum_{j\neq i=1}^{n_u}\lambda_j f^k_{u_j}\quad k\geq 0.
\end{align*}
Since, by definition, it holds that:
\begin{align*}
d^{k+1}\Omega_{k+1}=d^{k+1}\begin{pmatrix}
\Omega_{k}^T&
\Delta\Omega_{k+1}^T
\end{pmatrix}^T=
d^{k+1}\begin{pmatrix}
\Omega_{k}^T&L_{f_{xw}^{k}}\left(\Delta\Omega_{k}\right)^T&L_{f_{u_1}^{k}}\left(\Delta\Omega_{k}\right)^T&\cdots&L_{f_{u_{n_u}}^{k}}\left(\Delta\Omega_{k}\right)^T
\end{pmatrix}^T,
\end{align*}
the matrix built by ORC-DF algorithm in the $k+1-$th iteration includes $m\left(1+n_u\right)^{k+1}$ dependent rows; the rows forming $d^{k+1}L_{f_{u_i}^k}\left(\Delta\Omega_k\right)$ can be written as a linear combination of the remaining rows. Indeed, by a property of Lie derivative \citep{isidori1995nonlinear} it holds that:
\begin{align*}
&L_{f^k_{u_i}}\left(\Delta\Omega_k\right)=d^k\Delta\Omega_kf^k_{u_i}=d^k\Delta\Omega_k\left(\sum_{j\neq i=1}^{n_u}\lambda_j f_{u_j}^k\right)=\sum_{j\neq i=1}^{n_u}\lambda_jd^k\Delta\Omega_kf^k_{u_j}=
\sum_{j\neq i=1}^{n_u}\lambda_jL_{f^k_{u_j}}\left(\Delta\Omega_k\right),
\end{align*}
so, using linearity of the derivative, the following linear combination has been obtained:
\begin{align}\label{lin_comb}
&d^{k+1}L_{f^k_{u_i}}\left(\Delta\Omega_k\right)=d^{k+1}\left(\sum_{j\neq i=0}^{n_u}\lambda_jL_{f^k_{u_j}}\left(\Delta\Omega_k\right)\right)=\sum_{j\neq i=0}^{n_u}\lambda_jd^{k+1}L_{f^k_{u_j}}\left(\Delta\Omega_k\right)\in m \left(1+n_u\right)^{k+1}\times n^k
\end{align}

Likewise, 
if there exists $1\leq i\leq n_u$ such that $h_{u_i}$ is linearly dependent on vector fields $h_{u_j}$ for $1\leq j\leq n_u,$ $j\neq i,$ the observability matrix built by ORC-DF includes $m\left(1+n_u\right)^{k+1}$ dependent rows in the $k+1-$th iteration, for $k\geq 0.$

Note that the real values $\lambda_j$ can be replaced by functions of unknown parameters, $\lambda_j=\lambda_j\left(\theta\right),$ as they are constant variables, so the equality \eqref{lin_comb} holds for this case as well.

\section{Implementation}\label{sec:implementation}

\subsection{The STRIKE-GOLDD software toolbox}

STRIKE-GOLDD (STRuctural Identifiability taKen as Extended-Generalized Observability using Lie Derivatives and Decomposition) is an open source MATLAB toolbox that analyses the identifiability, observability, and invertibility of nonlinear systems of the form (\ref{din}--\ref{out}). It is available at \href{https://sites.google.com/site/strikegolddtoolbox/}{https://sites.google.com/site/strikegolddtoolbox/} and \href{https://github.com/afvillaverde/strike-goldd/}{https://github.com/afvillaverde/strike-goldd/}. 
STRIKE-GOLDD versions up to v2.1.6 implemented the FISPO algorithm, 
including a number of additional features that go beyond the core instructions described in Algorithm \ref{fispoalg}, with the purpose of facilitating the analysis of large models.
Furthermore, they also allowed to indicate a given number of non-zero time derivatives of inputs, both known and unknown.

\subsection{Implementation of the ORC-DF algorithm}

We have released a new version of STRIKE-GOLDD (v2.2) that includes an implementation of ORC-DF (Algorithm \ref{orc-alg}) along with the already existing implementation of FISPO (Algorithm \ref{fispoalg}).
The algorithm is chosen with the newly introduced option {\ttfamily opts.affine} in the {\ttfamily options.m} file (set it to 1 for ORC-DF, and to 0 for FISPO).
The {\ttfamily ORC\_DF.m} function checks whether a model is indeed affine in the inputs and, if that is the case, converts it to the appropriate form $\Sigma_A$ (\ref{dinaffine}--\ref{outaffine}), storing it in a mat-file to avoid repeating this calculation in the future. Thus, the user only needs to enter the model once, using the same format for ORC-DF and FISPO.
New specific options for the ORC-DF algorithm include the possibility of setting a maximum number of iterations through the variable {\ttfamily opts.kmax}, limiting the computation time of each stage with {\ttfamily opts.tStage}, and using the MATLAB Parallel Toolbox. 

\subsection{Multiple experiments and piecewise constant inputs}

FISPO analyses the observability of a model for a single experiment with an infinitely differentiable (``smooth'') input. However, it is possible to use it to consider multiple experiments with possibly different inputs by applying it to a modified model: if we create as many replicates of the model states, inputs, and outputs as the number of experiments that we want to consider, we obtain a new model whose analysis for a single input has the same observability properties as the original model with multiple inputs \citep{villaverde2019input}. 
Until now, this feature was only available in the GenSSI 2.0 toolbox \citep{ligon2018genssi}.
We have included the possibility of carrying out this multi-experiment analysis automatically in the new version of STRIKE-GOLDD, by setting the option {\ttfamily opts.multiexp=1}. The number of experiments can be chosen with {\ttfamily opts.numexp}.

\section{Computational results and discussion}\label{sec:results}

We have applied the ORC-DF and FISPO algorithms to a set of illustrative case studies from different areas of science and technology, ranging from civil engineering to different biological disciplines. They are  listed in Table \ref{tab:modresults}, along with the computation times of the algorithms.

\begin{table}[H]
\centering
\resizebox{\textwidth}{!}{
	\begin{tabular}{cccccc|ccc|ccc}
		\toprule
		\textbf{Model} & \textbf{Section} & \textbf{Reference} & \textbf{$\# \theta$} & \textbf{$\# u$}  & \textbf{$\# w$}  & \multicolumn{6}{c}{\textbf{Computation time [s]}}\\
		&&&&&&\multicolumn{3}{c}{\textbf{FISPO}}&\multicolumn{3}{c}{\textbf{ORC-DF}}\\
		&&&&&&$k=0$&$k=1$&$k=5$&$k=0$&$k=1$&$k=5$\\
		\midrule 
		C2M   & \ref{sec:c2m}    & \citep{villaverde2019full}      & 4 & 1 & 0 &      &$0.47$&     &        &$0.41$&   \\
		Bolie & \ref{sec:bolie}  & \citep{Bolie}		  	       & 5 & 1 & 0 &      &$1.42$&     &        &$0.59$&   \\
		2DOF  & \ref{sec:2dof}   & \citep{maes2019observability}   & 3 & 1 & 1 & $0.51$&$1.26$& $5.68$&$0.80$& $0.89$ &$1.54$\\
		HIV   & \ref{subsec:hiv} & \citep{miao2011identifiability} & 5 & 1 & 0 &      & $0.42$&     &        &$0.44$&     \\
		      &                  &                                 & 5 & 0 & 1 &$0.42$& $0.43$&$57.8$& $1.14$& $1.29$ & $47.1$\\    
		TS    & \ref{subsec:ts}  & \citep{lugagne2017balancing}    & 10& 2 & 0 &      & $99.9$ &     &        & N/A  &     \\
		      &                  &                                 & 6 & 0 & 2 & $1.47$ & $36.2$ & $>10^4$&        & N/A  &     \\
		JAK-STAT&\ref{subsec:jakstat}&\citep{bachmann2011division} & 26& 5 & 0 &      & Table \ref{tabla_jakstat} &  & & Table \ref{tabla_jakstat}  & \\
		\bottomrule
	\end{tabular}}
	\caption{\label{tab:modresults}Computation times of the two algorithms for the models analysed in this study.
	The computation times of case studies with unknown inputs depend on the highest order of the derivatives of the unknown inputs that are assumed to be non-zero, $k$. Three different cases are shown for those models: $k=\{0,1,5\}$.
	For models without unknown inputs this setting does not apply.
	Cases in which an algorithm cannot be applied 
	are labeled as N/A.
	Results were obtained on a personal computer with 16 GB RAM and processor Intel(R) Core(TM) i7-8550U 1.80 GHz.}
\end{table}

\subsection{An identifiable and observable model with known input: ``C2M''}\label{sec:c2m}

Our first case study is a deceivingly simple compartmental model \citep{villaverde2019input},
\begin{align*} 
&\begin{cases}
&\dot{x}_1(t)=-\left(k_{1e}+k_{12}\right)x_1(t)+k_{21}x_2(t)+bu(t)\\
&\dot{x}_2(t)=k_{12}x_1(t)-k_{21}x_2(t)\\
&y(t)=x_1(t)
\end{cases}
\end{align*}
where each state $x_i$ $\left(i=1,2\right)$ corresponds to a compartment, and $\theta=\left(k_{1e},k_{12},k_{21},b\right)$ is the unknown parameter vector. The augmented state vector is $\tilde x=\left(x_1,x_2,k_{1e},k_{12},k_{21},b\right),$ with extended dynamics given by:
\begin{align*} &f\left( x(t),\theta,u(t)\right)=f_{xw}\left(x(t),\theta,u(t)\right)+f_u\left(x(t),\theta,u(t)\right)u(t)=\begin{pmatrix}\dot{x}_1(t)&\dot{x}_2(t)&0&0&0&0\end{pmatrix}^T,
\end{align*}
with the following vector fields for the affine-in-inputs formulation (\ref{dynafin}):
\begin{align*}&f_{xw}\left(x(t),\theta,u(t)\right)=\begin{pmatrix}
-\left(k_{1e}+k_{12}\right)x_1(t)+k_{21}x_2(t)&k_{12}x_1(t)-k_{21}x_2(t)&0&0&0&0
\end{pmatrix}^T,\\
&f_u\left(x(t),\theta,u(t)\right)=\begin{pmatrix}
b&0&0&0&0&0
\end{pmatrix}^T.
\end{align*}
In addition, the output is given by the function:
\begin{align*}
&y(t)=h\left(x(t),\theta,u(t)\right)=h_{xw}\left(x(t),\theta,u(t)\right)=x_1(t).
\end{align*}

Due to its reduced size, this model is well suited for illustrating the differences between the procedures followed by the ORC-DF and FISPO algorithms. For this purpose we derive the equations of the extended Lie derivatives calculated by each algorithm in \ref{app-c2m}, where we also discuss the implications for the analyses.

\begin{figure}[H]
\centering
\includegraphics[width=\textwidth]{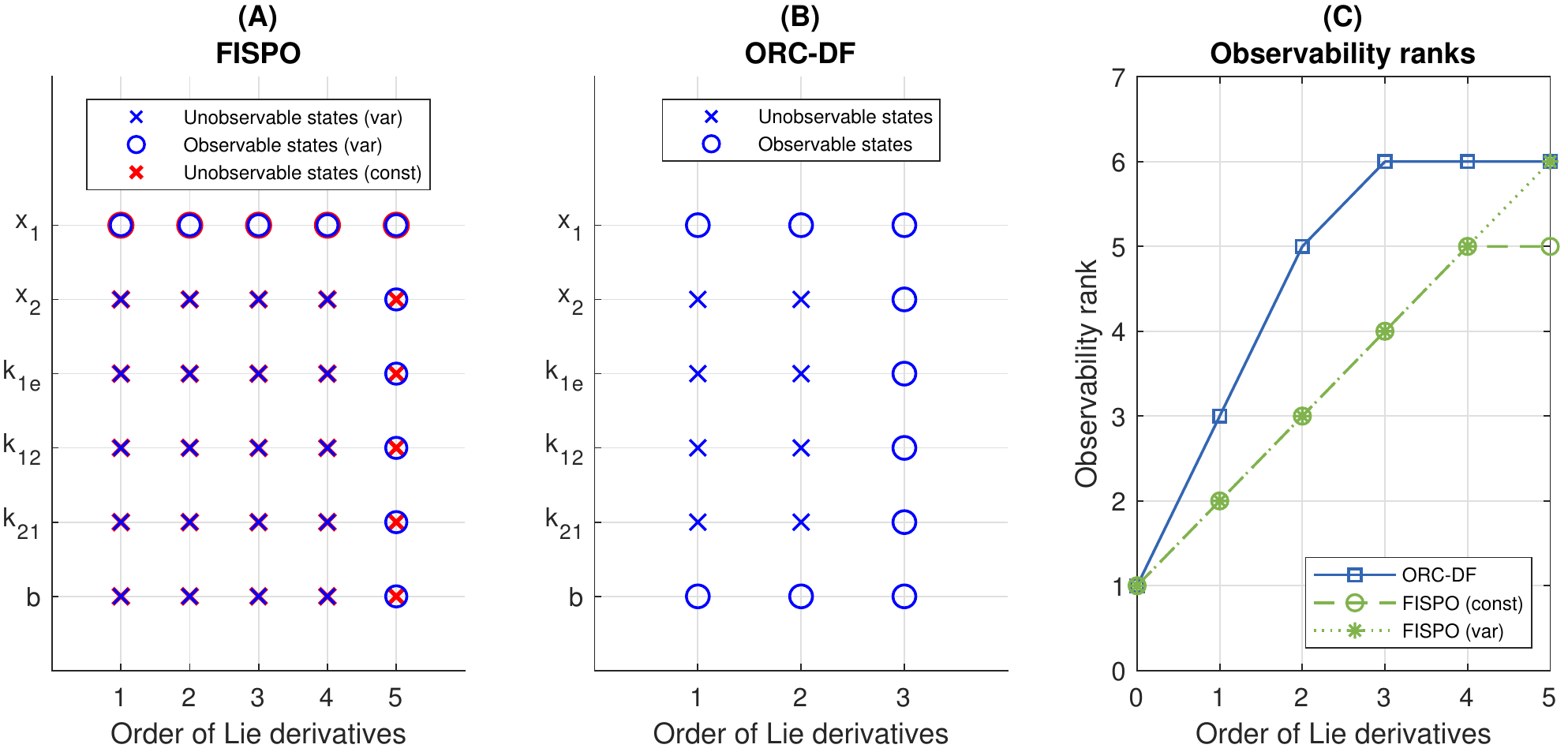}
	\caption{Analysis of the C2M model with the FISPO and ORC-DF algorithm. For FISPO two cases are considered: $\dot u = 0$, which is labeled as `(const)', and $\dot u \neq 0$, which is labeled as `(var)'.
	(A,B): Results of the FISPO and ORC-DF algorithms, respectively: the panel shows the states classified as observable or unobservable as a function of the number of Lie derivatives calculated by each algorithm. (C): Observability rank obtained by each algorithm as a function of the number of Lie derivatives. The full rank is equal to the number of states, i.e. six.}
	\label{c2m}
\end{figure}

This model is classified as observable and identifiable by ORC-DF after three iterations. The result yielded by FISPO depends on the number of input derivatives assumed to be zero: the unmeasured variables are classified as unobservable with a constant input, while they  become observable in the fifth iteration if the input is any non-constant analytical function. The variables classified as observable by both algorithms at each iteration are illustrated in Fig.~\ref{c2m}.A--B. Fig.~\ref{c2m}.C shows the ranks of the matrices built by both algorithms in each iteration. As can be seen, the observability matrix constructed by ORC-DF reaches full rank after considering Lie derivatives up to order three. The matrix built by FISPO stagnates from the fourth iteration onward with a constant input, while it reaches full rank after five iterations with a non-constant input.

\subsection{A non-identifiable, non-observable model with known inputs: ``Bolie''}\label{sec:bolie}

Our second example is a model with similarities to the previous one, given by \citep{Bolie}:
\begin{align*} 
&\begin{cases}
&\dot{q}_1(t)=p_1q_1(t)-p_2q_2(t)+u(t)\\
&\dot{q}_2(t)=p_4q_1(t)+p_3q_2(t)\\
&y(t)=\dfrac{1}{V_p}q_1(t)
\end{cases}
\end{align*}
where $x=\left(q_1,q_2\right)$ is the states vector, $\theta=\left(p_1,p_2,p_3,p_4,V_p\right)$ are the unknown parameters, and $u(t)$ is a measured input. The output is a function of the state $q_1$ and the unknown parameter $V_p\colon$
\begin{align*} 
h\left(x(t),\theta,u(t)\right)=h_{xw}\left(x(t),\theta, u(t)\right)=\frac{1}{V_p}q_1(t),
\end{align*}
so, in this case, there are no directly measured states or parameters.

The augmented state vector is $\tilde x=\left(q_1,q_2,p_{1},p_{2},p_{3},p_4,V_p\right),$ and the extended dynamics:
\begin{align*}
&f\left(x(t),\theta,u(t)\right)=f_{xw}\left(x(t),\theta,u(t)\right)+f_u\left(x(t),\theta,u(t)\right)u(t)=\begin{pmatrix}\dot{q}_1(t)&\dot{q}_2(t)&0&0&0&0&0\end{pmatrix}^T, \end{align*}
can be separated into the vector fields:
\begin{align*} & f_{xw}\left( x(t),\theta ,u(t)\right)=\begin{pmatrix}
p_1q_1(t)-p_2q_2(t)&p_4q_1(t)+p_3q_2(t)&0&0&0&0&0
\end{pmatrix}^T,\\
&f_u\left( x(t),\theta,u(t)\right)=\begin{pmatrix}
1&0&0&0&0&0&0
\end{pmatrix}^T.
\end{align*}

\begin{figure}[H]
	\centering 
	\includegraphics[width=\textwidth]{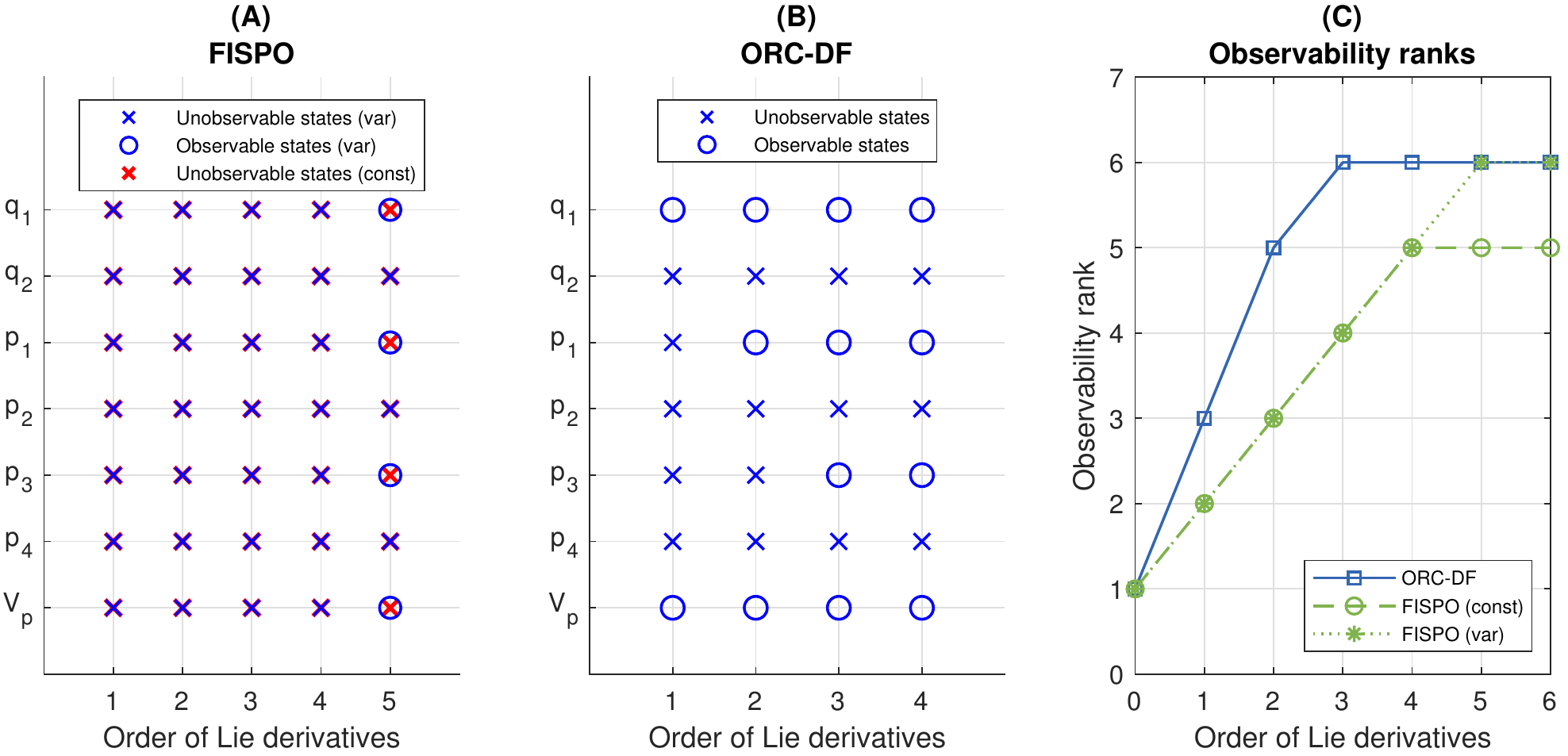}
	\caption{Analysis of the Bolie model with the FISPO and ORC-DF algorithm. For FISPO two cases are considered: $\dot u = 0$, which is labeled as `(const)', and $\dot u \neq 0$, which is labeled as `(var)'. (A,B). Results of the FISPO and ORC-DF algorithms, respectively: the panel shows the states classified as observable or unobservable as a function of the number of Lie derivatives calculated by each algorithm. (C) Observability rank obtained by each algorithm as a function of the number of Lie derivatives. The full rank is equal to the number of states, i.e. seven.}
	\label{rank_bol}
\end{figure}

The model is classified as non-identifiable and non-observable by FISPO and ORC-DC, as shown in Fig.~\ref{rank_bol}. 
A detailed analysis of the calculations performed by both algorithms is provided in \ref{app-bolie}.

\subsection{A model with known and unknown inputs: ``2DOF''}\label{sec:2dof}

We consider now an affine-in-the-inputs model with a known and an unknown input, proposed by \cite{maes2019observability}. It describes the behaviour of a mechanical system consisting of two masses connected by a spring. In the form (\ref{din}--\ref{out}), its dynamics and output functions are given by:
\begin{align*}
&    f\left(x(t),\theta,u(t),w(t)\right)=\begin{pmatrix}
dx_1(t)\\
dx_2(t)\\
\frac{1}{m_1}\left[-\left(k_1+\delta k_1x_1(t)\right)x_1(t)+k_2\left(x_2(t)-x_1(t)\right)-c_1dx_1(t)+c_2\left(dx_2(t)-dx_1(t)\right)+F_{1}(t)\right]\\
\frac{1}{m_2}\left[k_2\left(x_1(t)-x_2(t)\right)+c_2\left(dx_1(t)-dx_2(t)\right)+F_{2}(t)\right]
\end{pmatrix}
\end{align*}
\begin{align*}
&    h\left(x(t),\theta,u(t),w(t)\right)=h_{xw}\left(x(t),\theta,u(t),w(t)\right)=\begin{pmatrix}
x_1(t)&\frac{1}{m_2}\left[k_2\left(x_1(t)-x_2(t)\right)+c_2\left(dx_1(t)-dx_2(t)\right)+F_{2}(t)\right]
\end{pmatrix}^T.
\end{align*}

The state vector is $x=\left(x_1,x_2,dx_1,dx_2\right)$ and the unknown parameters are $\theta=\left(k_1,\delta k_1,m_2\right).$ Two external forces act on the system as inputs, one of known magnitude, $u(t)=F_1(t),$ and another of unknown value, $w(t)=F_2(t).$ The remaining parameters $k_2,$ $m_1,$ $c_1$ and $c_2$ are known.

Since there is an unknown input acting on the model, it is necessary to include its time derivatives in the extended states vector. The $0-$augmented state is $x^0=\left( x_1,x_2,dx_1,dx_2,k_1,\delta, k_1,m_2,w\right)$, which follows the dynamics:
\begin{align*}
&\dot{x}^0(t)=f^0\left(x^0(t),u(t),\dot{w}(t)\right)=f^0_{xw}(x^0(t),\dot{w}(t))+f^0_u\left( \tilde x(t) \right)u(t)=\begin{pmatrix} f\left(x(t),\theta,u(t),w(t)\right)^T&0&0&0&\dot{w}(t)\end{pmatrix}^T,
\end{align*}
where the contribution of the known input is:
\begin{align*}
f_u^0\left(\tilde x(t)\right)=\begin{pmatrix}
0&0&\frac{1}{m_1}&0&0&0&0&0
\end{pmatrix}^T.
\end{align*}

\begin{figure}[H]
	\centering
	\includegraphics[width=\textwidth]{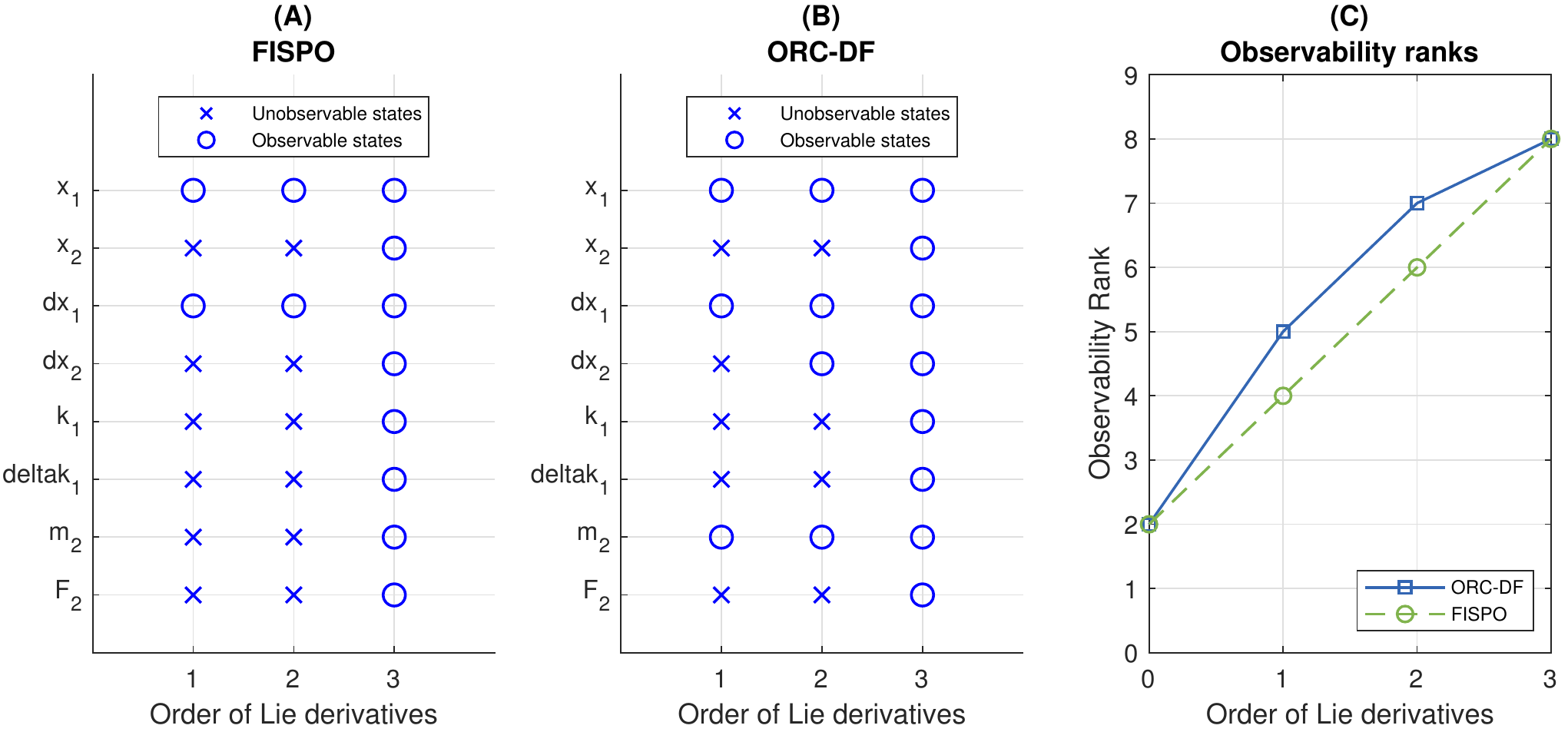}
	\caption{Analysis of the 2DOF model. (A,B). Results of the FISPO and ORC-DF algorithms, respectively, with $\dot{w}=0.$ The panels show the states classified as observable or unobservable as a function of the number of Lie derivatives calculated by each algorithm. (C) Observability rank obtained by each algorithm as a function of the number of Lie derivatives. The full rank is equal to the number of states, i.e. eight. 
	}
	\label{order_comp}
\end{figure}

First we consider the case in which the unknown disturbance $w(t)$ is assumed constant, $\dot w(t)=0$. The results are shown in Fig.~\ref{order_comp}. After calculating three Lie derivatives both algorithms conclude that the system is identifiable, observable and invertible. It should be noted that for this model FISPO always leads to the result shown in Fig.~\ref{order_comp}.A, regardless of the number of known input derivatives assumed to be non-zero.

\begin{figure}[H]
	\centering
	\includegraphics[width=\textwidth]{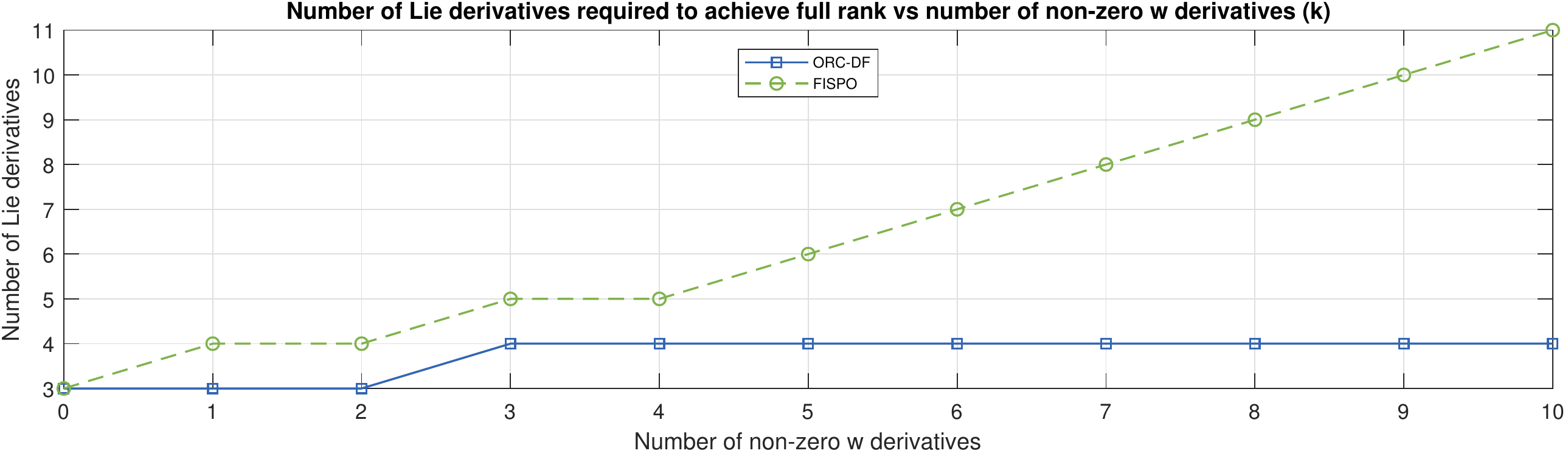}
	\caption{Number of Lie derivatives needed for building the observability matrices of the 2DOF model as a function of the number of derivatives of the unknown input $w(t)$ assumed to be non-zero.
	}
	\label{2dof_vs_k}
\end{figure}

Next, we consider a time-varying unknown input, assuming that $w^{\left.k\right)}=0$ for some $k>1.$ 
For this case, the model is again classified as fully observable by both algorithms. However, the paths that they follow to reach that conclusion are different.
The number of Lie derivatives required by FISPO to classify the model as observable increases as $k$ grows, due to the number of states in each stage also increasing without reaching full rank. In contrast, ORC-DF ends at most in four iterations, regardless of the value of $k\geq 1.$ This situation is illustrated in Fig.~\ref{2dof_vs_k}, which shows the number of derivatives required by each algorithm to achieve a result has been represented for $0\leq k\leq 10.$ 

A difference between the procedures carried out by both algorithms for this model is that -- similarly to the case of parameter $b$ in the C2M example, mentioned in \ref{app-c2m} -- the expressions obtained by ORC-DF determine that the unknown parameter $m_2$ can be calculated directly from the measurements since the first iteration as:
\begin{align*} 
&L_{f^0_u}h\left(x(t),\theta,u(t),w(t)\right)=\frac{c_2}{m_1m_2}.
\end{align*}
In the case of FISPO, in contrast, the parameter $m_2$ is the last to be classified as identifiable
, which happens at the same time in which the entry $w(t)$ is classified as invertible for a sufficiently large $k$ $(k\geq 5)$. 

\subsection{A model with a known or unknown input: ``HIV''}\label{subsec:hiv}

Next we consider a model of HIV dynamics in the human body given by \citep{miao2011identifiability}:
\begin{align*}
    \begin{cases}
        &\dot{T}_U(t)=\lambda-\rho T_U(t)-\eta(t)T_U(t)V(t)\\
        &\dot{T}_I(t)=\eta(t)T_U(t)V(t)-\delta T_I(t)\\
        &\dot{V}(t)=N\delta T_I(t)-cV(t)\\
        &y_1(t)=V(t)\\
        &y_2(t)=T_I(t)+T_U(t)
    \end{cases}
\end{align*}
where the states are $x=\left(T_U,T_I,V\right),$ the unknown parameters vector is $\theta=\left(\lambda, \rho,\delta,N,c\right),$ and $\eta(t)$ is a time-varying input, the infection rate. 

As was established in Section \ref{fispo=orc}, if $\eta(t)$ is unknown, ORC-DF and FISPO become the same algorithm (leaving aside implementation details). If $\eta(t)$ is known and time-varying, they differ.

This model was analysed with FISPO by \cite{villaverde2019full} considering two possibilities, i.e. $\eta(t)$ known and unknown. In both cases the model is classified as observable and identifiable by FISPO. In the latter case, the number of Lie derivatives necessary to achieve this conclusion grows with the number of derivatives of $\eta(t)$ assumed to be non-zero, as happened with the 2DOF model analysed in Section \ref{sec:2dof}.

It is possible to analyse the HIV model with the ORC-DF algorithm, since it is affine in inputs. With the infection rate considered known, i.e. $u(t)=\eta(t)$, the functions of the affine formulation (\ref{input_dyn_am}--\ref{input_dyn_am2}) are written as:
\begin{align*}
    &f_{xw}\left(x(t),\theta\right)=\begin{pmatrix}
    \lambda-\rho T_U(t)&-\delta T_I(t) & N\delta T_I(t)-cV(t)
    \end{pmatrix}^T\\
     &f_{u}\left(x(t),\theta\right)=\begin{pmatrix}
    -T_U(t)V(t)&T_U(t)V(t) & 0
    \end{pmatrix}^T\\
    &h_{xw}\left(x(t),\theta\right)=h\left(x(t),\theta,w(t)\right)=\begin{pmatrix}
    V(t)&T_I(t)+T_U(t)
    \end{pmatrix}^T
\end{align*}
while if it is unknown, $w(t)=\eta(t)$, we have:
\begin{align*}
    &f_{xw}\left(x(t),\theta,w(t)\right)=f\left(x(t),\theta,w(t)\right)\\
    &h_{xw}\left(x(t),\theta,w(t)\right)=h\left(x(t),\theta,w(t)\right)
\end{align*}

\begin{figure}[H]
	\centering 
	\includegraphics[width=\textwidth]{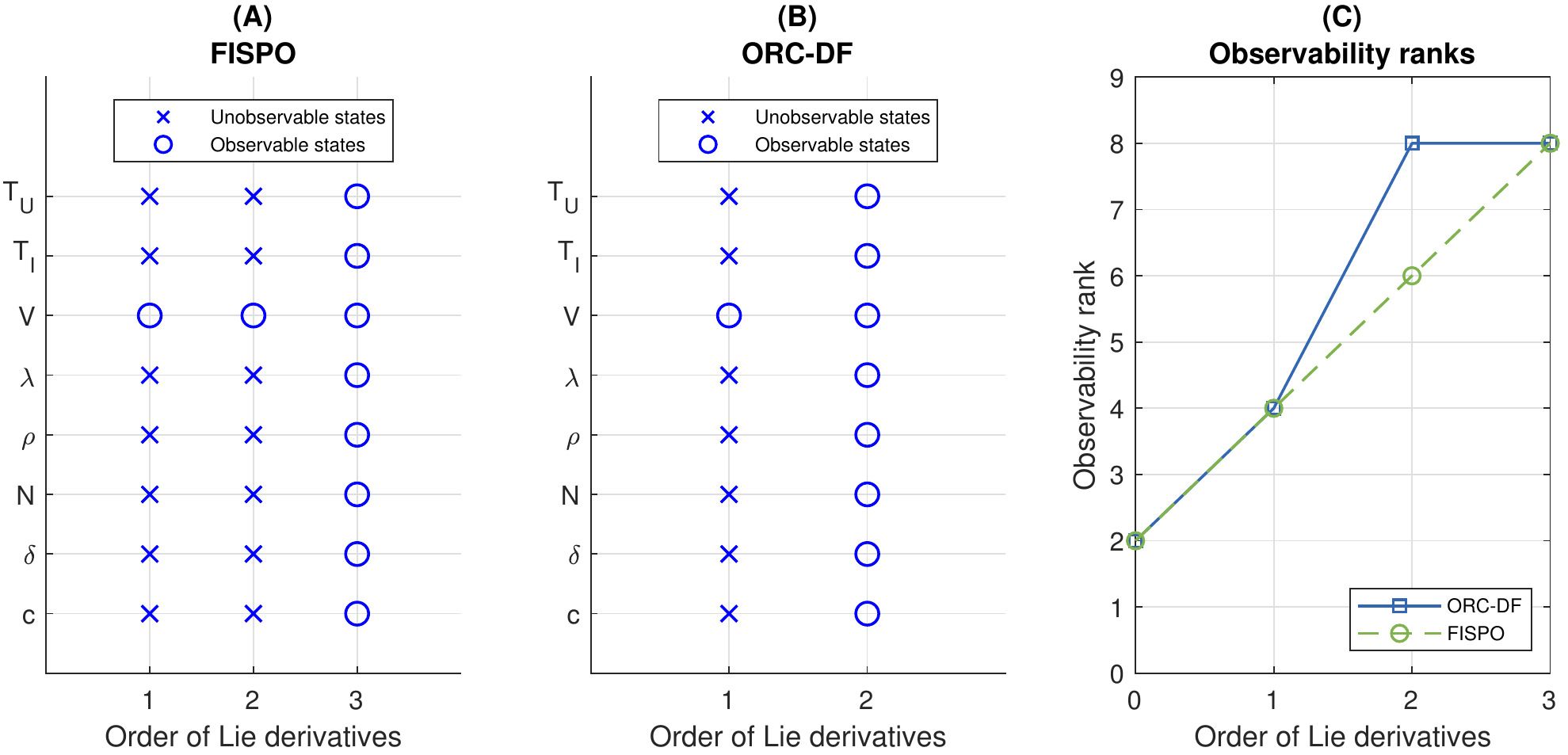}
	\caption{Analysis of the HIV model with the FISPO and ORC-DF algorithm, with the input $\eta(t)$ considered known. (A,B). Results of the FISPO and ORC-DF algorithms, respectively: the panel shows the states classified as observable or unobservable as a function of the number of Lie derivatives calculated by each algorithm. (C) Observability rank obtained by each algorithm as a function of the number of Lie derivatives. The full rank is equal to the number of states, i.e. eight.}
	\label{rank_hiv}
\end{figure}

If the infection rate is considered known, both algorithms classify the model as observable and identifiable, regardless of the number of input derivatives assumed non-zero by FISPO. Fig. \ref{rank_hiv} illustrates this fact.

The observability matrix calculated by ORC-DF has full rank after including Lie derivatives up to second order, so the number of its rows is
$
    \sum_{i=0}^{2}m\left(1+n_u\right)^{i+1}=\sum_{i=0}^{2}2^{i+2}=26
$ (actually, $14$ rows after excluding dependent rows arising from the equality $\left.h_u=0\right)$ while the matrix constructed by FISPO needs to include Lie derivatives up to order three to achieve full rank, so it has $m(k+1)=8$ rows. 
This is an example of a model for both algorithms perform similarly; although FISPO needs to calculate one more Lie derivative to classify the system as observable, ORC-DF calculates ranks of matrices of greater dimension, resulting in  similar computational cost of the calculations involved in each algorithm. As can be seen in Fig.\ref{rank_hiv}.C, the ranks of both observability matrices coincide up to the first iteration, as a consequence of:
\begin{align*}
    &h_u\left(x(t),\theta\right)=L_{f_u}h\left(x(t),\theta\right)=\begin{pmatrix}
    0&0
    \end{pmatrix}^T.
\end{align*}

\subsection{A genetic toggle switch with two inputs: ``TS''}\label{subsec:ts}

Let us now consider the following model of a genetic toggle switch \citep{lugagne2017balancing}:
\begin{align*}
&\begin{cases}
&\dot{x}_1(t)=k_{01}+\dfrac{k_1}{1+\left(x_2(t)/\left(1+\left(aTc(t)/\theta_{aTc}\right)^{\eta_{aTc}}\right)\right)^{\eta_{TetR}}}-x_1(t)\\
&\dot{x}_2(t)=k_{02}+\dfrac{k_2}{1+\left(x_1(t)/\left(1+\left(IPTG(t)/\theta_{IPTG}\right)^{\eta_{IPTG}}\right)\right)^{\eta_{LacI}}}-x_2(t)\\
&y_1(t)=x_1(t)\\
&y_2(t)=x_2(t)
\end{cases}
\end{align*}
where $x=\left(x_1,x_2\right)$ is the state vector and the inputs are $aTc(t)$ and $IPTG(t).$ The remaining variables are unknown parameters.

This model is an example that cannot be analysed by ORC-DF algorithm, since it is not affine in inputs.
It was analysed with FISPO in \citep{villaverde2019full}, considering both measured and unmeasured inputs. If both inputs are known, FISPO classifies the model as structurally identifiable, as long as neither input is constant. If the inputs are unknown FISPO concludes that some parameters become unidentifiable. For more details we refer the reader to  \citep{villaverde2019full}. 

\subsection{A signaling pathway with five known inputs: ``JAK-STAT''}\label{subsec:jakstat}

To show the computational limitations of the two algorithms, we analyse here a model that pushes them to their limits. It is a classic model of the JAK-STAT signaling pathway presented by \cite{bachmann2011division}, which has $25$ states, $26$ unknown parameters and $5$ inputs. The output consists on $15$ measured functions of the model variables that depend only on one of the external signals $u_i,$ which is not involved in system dynamics, that is,
\begin{align}\label{din1}
   &f_{u_i}\left(x(t),\theta,u(t)\right)=0_{25\times 1}\\\label{out1}
    &h_{u_j}\left(x(t),\theta,u(t)\right)=0_{15\times 1},\quad 1\leq j\leq 5,\quad j\neq i
\end{align}
The model equations are provided in \ref{app-jakstat}.

This model was analysed with FISPO in \citep{afvjakstat2019}, concluding that all its parameters are structurally identifiable but two of its 25 states are non-observable. The calculations are computationally expensive, requiring the use of procedures supported in STRIKE-GOLDD -- such as model decomposition or successive executions after removing parameters previously classified as identifiable -- in order to reach the conclusion. 
Thus, the model was first analysed after setting the maximum computation time of each Lie derivative to 100 seconds, which allowed FISPO to calculate $5$ Lie derivatives and to classify 17 parameters and 4 states as observable. Next, the 17 parameters were specified as previously classified in the FISPO options, thus removing them from further consideration and decreasing the size of the problem, and the model was decomposed. The post-decomposition analysis classified 5 additional parameters as identifiable. After removing them, it was possible to analyse the remainder of the model and reach the aforementioned conclusion.

With ORC-DF we did not manage to analyse the model due to computational limitations (specifically, insufficient memory). The different computational requirements of ORC-DF and FISPO are shown in Table \eqref{tabla_jakstat}. 

\begin{table}[H]
\centering
\resizebox{\textwidth}{!}{
	\begin{tabular}{l|ccccc|ccccc}
		\toprule
        &\multicolumn{5}{c|}{\textbf{ORC-DF}}&\multicolumn{5}{c}{\textbf{FISPO}}\\
        \midrule
        \textbf{Iteration}&1&2&3&4&5&1&2&3&4&5\\
        \textbf{Number of rows}&$180$ &$930$&$4680$&$23430$&$117180$&$30$&$45$&$60$&$75$&$90$\\
        \textbf{Rank}&$24$&$35$&$40$&$42$&$44$&$20$&$28$&$34$&$39$&$43$\\
        \textbf{Rank computation time [s]}&$0.54$&$2.18$&$15.36$& $170.95$&$ 3600.54$&$0.40$&$0.76$&$3.46$&$59.55$&$ 362.78$\\
        \textbf{Observable variables}&$14$&$21$&$29$&$29$&$29$&$8$&$12$&$15$&$15$&$21$\\
	    \bottomrule
	\end{tabular}}
	\caption{Results and computation times of ORC-DF and FISPO for the JAK-STAT model.}
	\label{tabla_jakstat}
\end{table}

Table \eqref{tabla_jakstat} shows that the number of rows of the matrix built by ORC-DF grows rapidly at each iteration (even though the implementation removes null rows arising from dependencies in (\ref{din1}--\ref{out1}), which is why the number of rows of the ORC-DF matrix does not match the number given in Section \ref{lab_section}). Although this matrix leads to higher ranks than the one built by FISPO, especially at the beginning of the execution (i.e. with few Lie derivatives), the difference decreases soon and the rank of the two matrices is similar despite the big difference in the number of rows. With 5 Lie derivatives ORC-DF labels $27$ model variables as observable ($9$ states and $18$ parameters), including those classified as observable by FISPO (21: 4 states and 17 parameters). However, the computation time of ORC-DF at that point is roughly ten times higher than FISPO, and memory requirements impede further progress with this algorithm.

\section{Conclusion}\label{sec:conclusion}

In this paper we have analysed two recent algorithms for observability analysis of nonlinear systems with known and/or unknown inputs, which we refer to as ORC-DF and FISPO. 
Our analyses have revealed the key similarities and differences between them. The main conclusions can be summarized as follows.

First, we have proven theoretically that for models without known inputs both algorithms are basically equivalent, since they calculate the rank of the same observability matrix.
In contrast, for models with known inputs -- e.g. for controlled systems -- the two algorithms differ, since they build different observability matrices. Specifically, the number of rows of the matrix built by ORC-DF increases more at each iteration than the one built by FISPO. We have shown that this increased growth is often an advantage of ORC-DF, since it makes it possible to reach full rank -- and, thus, to conclude that a model is observable -- with less Lie derivatives, and hence less computational cost; an example was shown in Section \ref{sec:c2m}.
However, said increased growth is not always advantageous: as we have noted in Section \ref{lab_section}, it can also be detrimental to the efficiency of ORC-DF. The latter situation may happen when the structure of the model equations is such that the increase in problem dimension outweighs the increase in information resulting from the inclusion of a new Lie derivative; we provided an example in Section \ref{subsec:jakstat}. 

When applying FISPO to a model with unknown input(s), it is generally necessary to assume that the derivatives of the unknown input, $w^{k)}$, are zero for orders higher than a finite $k$. This is not a theoretical requirement -- and in fact, a counter-example that did not require this assumption was shown in \cite{villaverde2019full} -- but it is often necessary in practice in order to reach a conclusion in finite time. This was indeed the case for the models with unknown inputs that we analysed in this paper.

Another difference between both algorithms lies in the types of models and known inputs that they can analyse. In regard to model types, FISPO is applicable to a general class of nonlinear ODE models, while ORC-DF is applicable to a subclass of those models: the ones that are affine in the inputs. 
There is an additional, albeit subtle, difference between both algorithms in regard to the (known) inputs:
FISPO considers infinitely differentiable (``smooth'') functions, while ORC-DF considers piecewise constant inputs.
It should be noted however that an affine system that is observable for piecewise constant inputs is also observable for smooth inputs; therefore, ORC-DF can also establish the observability of (affine) models with continuous inputs. 
Finally, FISPO can analyse observability from multiple experiments by applying a model transformation as described in Section \ref{sec:implementation}.
We have implemented the possibility of carrying out this transformation automatically in a new version of the STRIKE-GOLDD toolbox (v2.2), thus enabling the FISPO algorithm to consider multiple experiments without requiring any manual transformation from the user. 
In STRIKE-GOLDD 2.2 we have also included an implementation of the ORC-DF algorithm, thus allowing the user to apply different algorithms with the same tool and model definition. It should be noted that the implementations of the FISPO and ORC-DF algorithms included in the STRIKE-GOLDD toolbox have a number of additional features that increase the efficiency of the core algorithms analysed here, as noted in Section \ref{sec:implementation}.

We used the new implementations in STRIKE-GOLDD 2.2 to benchmark the algorithms with several models taken from the literature. 
Our selection of case studies included both simple models, included to illustrate the inner workings of the algorithms in detail, and more complex models whose analysis is computationally challenging, which we used for pushing the algorithms to their limits. We also provided an example of a model that cannot be analysed with ORC-DF due to being not affine in the inputs. 

In conclusion, the theoretical and computational analyses presented here have informed about the differences between the ORC-DF and FISPO algorithms, showing that they represent complementary techniques for solving an often challenging problem, and clarifying when one may be preferred over the other. The release of a new version of the MATLAB toolbox STRIKE-GOLDD that includes implementations of both algorithms provides the convenience of performing different analyses with minimal intervention from the user.

\section*{Data and materials availability}

The methods and models used in this paper are available in the GitHub repository as part of release v2.2 of the STRIKE-GOLDD toolbox: \url{https://github.com/afvillaverde/strike-goldd}. 

\section*{Author contributions}

A.F.V. designed and supervised the research, N.M. implemented the software and performed the computational experiments, N.M. and A.F.V. analysed the algorithms and the results, N.M. and A.F.V. wrote the manuscript.

\section*{Acknowledgement}

This research was supported by the Spanish Ministry of Science, Innovation and Universities through the project SYNBIOCONTROL (ref. DPI2017-82896-C2-2-R).

\appendix

\section{Analysis of the Lie derivatives of the C2M case study}\label{app-c2m}

Let us first consider the FISPO algorithm and the model with non-constant input. As shown in Fig.~\ref{c2m}.C, in this case there are six independent extended Lie derivatives of the output:
\begin{align}
\label{ec-1}
&L_f^0h\left(x(t),\theta,u(t)\right)=y(t)=x_1\\
\label{ec2}
&L_fh\left(x(t),\theta,u(t)\right)=y'(t)=-\left(k_{1e}+k_{12}\right)x_1(t)+k_{21}x_2(t)+bu(t)\\
\label{ec3}
&L_f^2h\left(x(t),\theta,u(t)\right)=y''(t)=\left(\left(k_{1e}+k_{12}\right)^2+k_{21}k_{12}\right)x_1(t)-
\left(k_{1e}+k_{12}+k_{21}\right)k_{21}x_2(t)-\left(k_{1e}+k_{12}\right)bu(t)+b\dot{u}(t)\\
\nonumber 
&L_f^3h\left(x(t),\theta,u(t)\right)=y'''\left(t\right)=-\left[\left(k_{1e}+k_{12}\right)^3+k_{12}k_{21}\left(k_{1e}+k_{12}\right)+k_{12}k_{21}\left(k_{1e}+k_{12}+k_{21}\right)\right]x_1(t)-\left(k_{1e}+k_{12}\right)b\dot{u}(t)+\\
\label{ec-4}
&\left[\left(k_{1e}+k_{12}\right)^2+k_{12}k_{21}+k_{21}\left(k_{1e}+k_{12}+k_{21}\right)\right]k_{21}x_2(t)+\left[\left(k_{1e}+k_{12}\right)^2+k_{21}k_{12}\right]bu(t)+b\ddot{u}(t)
\\\nonumber 
&L^4_{f}h\left(x(t),\theta,u(t)\right)=y^{\left.4\right)}\left(t\right)=\left[\left(k_{1e}+k_{12}\right)^4+2k_{12}k_{21}\left(k_{1e}+k_{12}\right)^2+k_{12}k_{21}\left(k_{1e}+k_{12}+k_{21}\right)^2+k_{12}^2k_{21}^2\right]x_1-\\\nonumber
&\left[\left(k_{1e}+k_{12}\right)^3+2k_{12}k_{21}\left(k_{1e}+k_{12}\right)+2k_{12}k_{21}^2+k_{21}\left(k_{1e}+k_{12}\right)^2+k_{21}^2\left(k_{1e}+k_{12}+k_{21}\right)\right]k_{21}x_2-b\left(k_{1e}+k_{12}\right)\ddot{u}(t)-\\\label{ec-5}
&\left[\left(k_{1e}+k_{12}\right)^3+2k_{12}k_{21}\left(k_{1e}+k_{12}\right)+k_{12}k_{21}^2\right]bu+\left[\left(k_{1e}+k_{12}\right)^2+k_{21}k_{12}\right]b\dot{u}(t)+b\dddot{u}(t)
\\
\nonumber 
&L^5_{f}h\left(x(t),\theta,u(t)\right)=y^{\left.5\right)}\left(t\right)=-\left[\left(k_{1e}+k_{12}\right)^5+3k_{12}k_{21}\left(k_{1e}+k_{12}\right)^3+3k_{12}^2k_{21}^2\left(k_{1e}+k_{12}\right)+k_{12}k_{21}^2\left(k_{1e}+k_{12}\right)^2+\right.\\\nonumber 
&\left.k_{12}k_{21}\left(k_{1e}+k_{12}\right)\left(k_{1e}+k_{12}+k_{21}\right)^2+k_{12}k_{21}^3\left(k_{1e}+k_{12}+k_{21}\right)+2k_{12}^2k_{21}^2\right]x_1(t)+\left[\left(k_{1e}+k_{12}\right)^2+k_{21}k_{12}\right]b\ddot{u}(t)+\\\nonumber 
&\left[\left(k_{1e}+k_{12}\right)^4+k_{21}\left(k_{1e}+k_{12}\right)^3+3k_{12}^2k_{21}^2\left(k_{1e}+k_{12}\right)+k_{21}\left(k_{1e}+k_{12}\right)^2\left(2k_{12}+k_{21}\right)+k_{12}k_{21}\left(k_{1e}+k_{12}+k_{21}\right)^2+\right.\\\nonumber
&\left.2k_{12}k_{21}^2\left(k_{1e}+k_{12}\right)+k_{21}^3\left(k_{1e}+k_{12}+k_{21}\right)+2k_{12}k_{21}^3+k_{12}^2k_{21}^2\right]k_{21}x_2-b\left(k_{1e}+k_{12}\right)\dddot{u}(t)+bu^{\left.4\right)}(t)+\\\nonumber
&\left[\left(k_{1e}+k_{12}\right)^4+2k_{12}k_{23}\left(k_{1e}+k_{12}\right)^2+k_{12}k_{21}\left(k_{1e}+k_{12}+k_{21}\right)^2+k_{12}^2k_{21}^2\right]bu(t)+\\
\label{ec-6}
&\left[\left(k_{1e}+k_{12}\right)^3+2k_{12}k_{21}\left(k_{1e}+k_{12}\right)+k_{12}k_{21}^2\right]b\dot{u}
\end{align}
Assuming for simplicity that $u^{\left.k\right)}=0$ for $k>1$, from the equations (\ref{ec-1}--\ref{ec-6}) we obtain:
\begin{align}\label{ec-red1}
&x_1(t)=y(t)\\
\label{ec-red2}
&bu(t)=y'(t)+\left(k_{1e}+k_{12}\right)y(t)-k_{21}x_2(t)\\
\label{ec-red3}   
&k_{21}^2x_2(t)=k_{12}k_{21}y(t)-\left(k_{1e}+k_{12}\right)y'(t)-y''(t)+b\dot{u}(t)\\
\label{ec-red4}
&k_{21}b\dot{u}(t)=y'''(t)+k_{1e}k_{21}y'(t)+\left(k_{1e}+k_{12}+k_{21}\right)y''(t)\\
\label{ec-red5}
&k_{1e}k_{21}y''(t)=-y^{\left.4\right)}(t)-\left(k_{1e}+k_{12}+k_{21}\right)y'''(t)\\
\label{ec-red6}
&\left(k_{1e}+k_{12}+k_{21}\right)\left(y'''^2(t)-y^{\left.4\right)}(t)y''(t)\right)=y^{\left.5\right)}(t)y''(t)-y^{\left.4\right)}(t)y'''(t)
\end{align}
Therefore, we can extract directly the following expressions:
\begin{align}
\label{ex-x1}
&x_1(t)=\phi_0\left(y,y',\dots,y^{\left.5\right)},u,\dot{u}\right)\\
\label{ex-123}
&k_{1e}+k_{12}+k_{21}=\phi_1\left(y,y',\dots,y^{\left.5\right)},u,\dot{u}\right)\\
\label{ex-13}
&k_{1e}k_{21}=\phi_2\left(y,y',\dots,y^{\left.5\right)},u,\dot{u}\right)\\
\label{ex-3b}
&k_{21}b=\phi_3\left(y,y',\dots,y^{\left.5\right)},u,\dot{u}\right)
\end{align}
where $\phi_i$ $\left(0\leq i\leq 3\right)$ are functions that depend only on the output, the input, and their time derivatives. It is possible to obtain similar input-output expressions for any variable involved in the model by determining the unique solution of the system  (\ref{ec-red1}--\ref{ec-red6}), which consists of six independent equations and six unknowns.

By inspecting equations (\ref{ec-red1}--\ref{ec-red6}) it is possible to explain the classification obtained by FISPO and shown in Fig.~\ref{c2m}.A (blue line), since it is not possible to obtain such an input-output expression for any unmeasured variable until calculating the fifth Lie derivative \eqref{ec-6}. (Note that this result requires excluding those states and parameters in the phase space for which the denominators in the input-output expressions vanish. However, since the system is analytical these states form a zero measurement subset).

Let us consider now the FISPO algorithm in the constant input case. From Fig.~\ref{c2m}.C it follows that the equations obtained (\ref{ec-1}--\ref{ec-6}) are dependent. The following system of equations is extracted from (\ref{ec-1}--\ref{ec-5}):
\begin{align}\label{ecu-red1}
&x_1(t)=y(t)\\
\label{ecu-red2}
&k_{21}x_2=y'(t)+\left(k_{1e}+k_{12}\right)y(t)-bu(t)\\
\label{ecu-red3}   
&k_{21}bu=y''(t)+k_{1e}k_{21}y(t)+\left(k_{1e}+k_{12}+k_{21}\right)y'(t)\\
\label{ecu-red4}
&k_{1e}k_{21}y'(t)=-y'''(t)-\left(k_{1e}+k_{12}+k_{21}\right)y''(t)\\
\label{ecu-red5}
&\left(k_{1e}+k_{12}+k_{21}\right)\left(y''^2(t)-y'''(t)y'(t)\right)=y'(t)y^{\left.4\right)}(t)-y''(t)y'''(t)
\end{align}
Using equations (\ref{ecu-red1}--\ref{ecu-red5}) it is possible to write the combinations $x_1,$ $k_{21}b,$ $k_{1e}+k_{12}+k_{21}$ y $k_{1e}k_{21}$ exclusively as functions of the input and output of the model. In this case the system of equations to be solved is:
\begin{align}
&x_1(t)=\phi_0\left(y,y',\dots,y^{\left.4\right)},u\right)\\
&k_{1e}+k_{12}+k_{21}=\phi_1\left(y,y',\dots,y^{\left.4\right)},u\right)\\
&k_{1e}k_{21}=\phi_2\left(y,y',\dots,y^{\left.4\right)},u\right)\\
&k_{21}b=\phi_3\left(y,y',\dots,y^{\left.4\right)},u\right)\\
&k_{21}^2\left(x_2(t)+x_1(t)\right)=k_{21}\left(y'(t)+\phi_1\left(y,y',\dots,y^{\left.4\right)},u\right)y(t)\right)-\phi_3\left(y,y',\dots,y^{\left.4\right)},u\right)u(t)
\end{align}
which has six unknowns and five independent equations, so it is not possible to write any of the parameters or the unknown state $x_2$ as a function of the input and output only. This scenario is shown in Fig.~\ref{c2m}.A (red line).

We consider now the ORC-DF algorithm. Instead of (\ref{ec-1}--\ref{ec-6}), it computes the following Lie derivatives:
\begin{align}
\label{ecorc-1}
&L_0(t)=h\left(x(t),\theta,u(t)\right)=x_1(t)\\
\label{ecorc-2}
&L_1(t)=L_{f_{xw}}h\left(x(t),\theta,u(t)\right)=-\left(k_{1e}+k_{12}\right)x_1(t)+k_{21}x_2(t)\\
\label{ecorc-3}
&L_2(t)=L_{f_u}h\left(x(t),\theta,u(t)\right)=b\\
\label{ecorc-4}
&L_3(t)=L^2_{f_{xw}}h\left(x(t),\theta,u(t)\right)=\left[\left(k_{1e}+k_{12}\right)^2+k_{21}k_{12}\right]x_1(t)-\left(k_{1e}+k_{12}+k_{21}\right)k_{21}x_2(t)\\
\label{ecorc-5}
&L_4(t)=L_{f_u}L_{f_{xw}}h\left(x(t),\theta,u(t)\right)=-b\left(k_{1e}+k_{12}\right)\\
\label{ecorc-6}
&L_5(t)=L^3_{f_{xw}}h\left(x(t),\theta,u(t)\right)=-\left[\left(k_{1e}+k_{12}\right)^3+2k_{21}k_{12}\left(k_{1e}+k_{12}\right)+k_{21}^2k_{12}\right]x_1(t)+\left[\left(k_{1e}+k_{12}\right)^2+k_{21}^2+\right.\\\nonumber 
& \left.k_{21}\left(k_{1e}+2k_{12}\right)\right]k_{21}x_2(t)\\
\label{ecorc-7}
&L_6(t)=L_{f_u}L^2_{f_{xw}}h\left(x(t),\theta,u(t)\right)=b\left[\left(k_{1e}+k_{12}\right)^2+k_{21}k_{12}\right]
\end{align}

The system (\ref{ecorc-1}--\ref{ecorc-7}) has a unique solution, in which one of the equations depends on the others. The solution, obtained from (\ref{ecorc-1}--\ref{ecorc-5}) and \eqref{ecorc-7}, is given as a function of Lie derivatives $L_i(t),$ $i\in\left\lbrace 0,1,2,3,4,6\right\rbrace,$ as follows:
  \begin{align*}
  &x_1(t)=L_0(t)\\
  &x_2(t)=\frac{\left(L_1(t)L_2(t)-L_0(t)L_4(t)\right)^2}{L_2(t)\left(L_0(t)L_6(t)-L_2(t)L_3(t)\right)+L_4(t)\left(L_1(t)L_2(t)-L_0(t)L_4(t)\right)}\\
  &k_{1e}=\frac{L_2(t)\left(L_3(t)L_4(t)-L_1(t)L_6(t)\right)}{L_2(t)\left(L_0(t)L_6(t)-L_2(t)L_3(t)\right)+L_4(t)\left(L_1(t)L_2(t)-L_0L_4(t)\right)}\\
  &k_{12}=\frac{\left(L_1(t)L_2(t)-L_0(t)L_4(t)\right)\left(L_2(t)L_6(t)-L_4^2(t)\right)}{L_2(t)\left(L_0(t)L_6(t)-L_2(t)L_3(t)\right)+L_4(t)\left(L_1(t)L_2(t)-L_0L_4(t)\right)}\\
  &k_{21}=\frac{L_2(t)\left(L_0(t)L_6(t)-L_2(t)L_3(t)\right)+L_4(t)\left(L_1(t)L_2(t)-L_0(t)L_4(t)\right)}{L_2(t)\left(L_1(t)L_2(t)-L_0(t)L_4(t)\right)}\\
  &b=L_2(t)
\end{align*}
Therefore, the ORC-DF algorithm classifies the C2M model as observable and identifiable after the third iteration. We note that equation \eqref{ecorc-3} implies that parameter $b$ can be calculated directly from the output.


\section{Analysis of the Lie derivatives of the Bolie case study}\label{app-bolie}

Assuming non-constant input, FISPO calculates six independent Lie derivatives of the output, as shown in Fig.~\ref{rank_bol}.C:
\begin{align}
\label{ecbol-fis-1}
&L^0_fh\left(x(t),\theta,u(t)\right)=y(t)=\frac{1}{V_p}q_1(t)\\
&L_fh\left(x(t),\theta,u(t)\right)=y'(t)=\frac{1}{V_p}\left(p_1q_1(t)-p_2q_2(t)\right)+\frac{1}{V_p}u(t)\\
&L^2_fh\left(x(t),\theta,u(t)\right)=y''(t)=\frac{1}{V_p}\left[\left(p_1^2-p_2p_4\right)q_1(t)-\left(p_1+p_3\right)p_2q_2(t)\right]+\frac{1}{V_p}\left(p_1u(t)+\dot{u}(t)\right)\\\nonumber
&L^3_fh\left(x(t),\theta,u(t)\right)=y'''(t)=\frac{1}{V_p}\left[p_1\left(p_1^2-p_2p_4\right)-p_2p_4\left(p_1+p_3\right)\right]q_1(t)+\frac{1}{V_p}\left[\left(p_1^2-p_2p_4\right)u(t)+p_1\dot{u}(t)\right]+\\
&\frac{1}{V_p}\ddot{u}(t)-\frac{1}{V_p}\left[p_1^2-p_2p_4+p_3\left(p_1+p_3\right)\right]p_2q_2(t)\\
\label{ecbol-fis-5}\nonumber
&L^4_fh\left(x(t),\theta,u(t)\right)=y^{\left.4\right)}(t)=\frac{1}{V_p}\left[\left(p_1^2-p_2p_4\right)^2-p_2p_4\left(p_1+p_3\right)^2\right]q_1(t)+\frac{1}{V_p}\left(p_1^3-2p_1p_2p_4-p_2p_3p_4\right)u(t)-\\
&\frac{1}{V_p}\left(p_1+p_3\right)\left(p_1^2-2p_2p_4+p_3^2\right)p_2q_2(t)+\frac{1}{V_p}\left[\left(p_1^2-p_2p_4\right)\dot{u}(t)+p_1\ddot{u}(t)+\dddot{u}(t)\right]\\
\label{ecbol-fis-6}\nonumber 
&L^5_fh\left(x(t),\theta,u(t)\right)=y^{\left.5\right)}(t)=\frac{1}{V_p}\left[p_1^2\left(p_1-p_2p_4\right)^2-p_2p_4\left(p_1+p_3\right)\left(2p_1^2+p_3^2-p_2p_4\right)\right]q_1(t)-\\\nonumber
&\frac{1}{V_p}\left[\left(p_1-p_2p_4\right)^2-p_2p_4\left(p_1+p_3\right)^2+p_3\left(p_1+p_3\right)\left(p_1^2-2p_2p_4+p_3^2\right)\right]p_2q_2(t)+\\\nonumber 
&\frac{1}{V_p}\left[\left(p_1^2-p_2p_4\right)^2-p_2p_4\left(p_1+p_3\right)^2\right]u(t)+\frac{1}{V_p}\left[p_1^3-2p_1p_2p_4-p_2p_3p_4\right]\dot{u}(t)+\\
&\frac{1}{V_p}\left(p_1^2-p_2p_4\right)\ddot{u}(t)+\frac{1}{V_p}\left(p_1\dddot{u}(t)+u^{\left.4\right)}(t)\right)
\end{align}

System (\ref{ecbol-fis-1}--\ref{ecbol-fis-6}) has no unique solution, since its composed by six equations and seven unknowns, but it is possible to obtain from it an input-output expression for the variables $q_1,$ $p_1,$ $p_3$ and $V_p,$ as it is shown in Fig. \ref{rank_bol}.

In the case of constant input, equations (\ref{ecbol-fis-1}--\ref{ecbol-fis-6}) become redundant, so there are only five independent Lie derivatives, as it is shown in Fig. \ref{rank_bol}. The combinations $p_1+p_3,$ $p_1p_3+p_2p_4$ are the only observable functions of the variables that can be extracted from (\ref{ecbol-fis-1}--\ref{ecbol-fis-5}). Since the rational combinations of these functions are insufficient to determine any of the parameters or states, all variables are non-observable, as shown in Fig.~\ref{rank_bol}.A (red line).

On the other hand, ORC-DF calculates the following Lie derivatives:
\begin{align}
\label{ecbol-1}
&h\left(x(t),\theta,u(t)\right)=\frac{1}{V_p}q_1(t)\\
\label{ecbol-2}
&L_{f_{xw}}h\left(x(t),\theta,u(t)\right)=\frac{1}{V_p}\left(p_1q_1(t)-p_2q_2(t)\right)\\
\label{ecbol-3}
&L_{f_u}h\left(x(t),\theta,u(t)\right)=\frac{1}{V_p}\\
\label{ecbol-4}
&L^2_{f_{xw}}h\left(x(t),\theta,u(t)\right)=\frac{1}{V_p}\left[\left(p_1^2-p_2p_4\right)q_1(t)-\left(p_1+p_3\right)p_2q_2(t)\right]\\
\label{ecbol-5}
&L_{f_u}L_{f_{xw}}h\left(x(t),\theta,u(t)\right)=\frac{1}{V_p}p_1\\
\label{ecbol-6}
&L^3_{f_{xw}}h\left(x(t),\theta,u(t)\right)=\frac{1}{V_p}\left[\left(p_1\left(p_1^2-p_2p_4\right)-\left(p_1+p_3\right)p_2p_4\right]q_1(t)-\left[\left(p_1^2-p_2p_4\right)+p_3\left(p_1+p_3\right)\right)p_2q_2(t)\right]\\
\label{ecbol-7}
&L_{f_u}L^2_{f_{xw}}h\left(x(t),\theta,u(t)\right)=\frac{1}{V_p}\left(p_1^2-p_2p_4\right)
\end{align}

From equations \eqref{ecbol-1}, \eqref{ecbol-3} and \eqref{ecbol-5} it is easy to obtain the state $q_1$ and the parameters $V_p$ and $p_1$ uniquely from the measurements, using Lie derivatives up to order two. This property is not fulfilled by the system formed by (\ref{ecbol-fis-1}--\ref{ecbol-fis-6}); Fig.~\ref{rank_bol}.A shows that the aforementioned variables are classified as observable by FISPO only after considering fifth order derivatives. Using the input-output expressions of $q_1,$ $V_p$ and $p_1$ extracted from \eqref{ecbol-1}, \eqref{ecbol-3} and \eqref{ecbol-5}, in conjunction with equations \eqref{ecbol-2}, \eqref{ecbol-4} and \eqref{ecbol-7}, it is also possible to determine parameter $p_3$ as a function of the Lie derivatives of the output. However, Fig.~\ref{rank_bol}.C shows that system (\ref{ecbol-1}--\ref{ecbol-7}) contains one redundant equation, so it is not possible to determine uniquely an input-output expression of the remaining unmeasured states
. It can also be noted that any rational combination of the observable variables with the functions $p_2q_2$ and $p_2p_4$ is also observable.

\section{Equations of the JAK-STAT model}\label{app-jakstat}

The dynamics of the JAK-STAT model analysed in Section \ref{subsec:jakstat} is given by:
\begin{equation*}\label{eqjak}
\begin{array}{lll}
\dot{x}_1& = &x_{2345}x_8\theta_{11}/\theta_{26} - k_5 x_1 \theta_{10}/M_1 ,\\ 
\dot{x}_2& = &k_5 x_1 \theta_{10}/M_1 - x_2 \theta_{7}/M_1 - x_2 x_8 \theta_{11} / \theta_{26} - 3 x_2 \theta_{7}/((\theta_{8}  x_{6} + 1) M_1),\\ 
\dot{x}_3& = & \theta_{7}x_2/M_1 - \theta_{11} x_8 x_{3} / \theta_{26}- 3 \theta_{7} x_{3}/((\theta_{8} x_{6} + 1) M_1) ,\\ 
\dot{x}_4& = &3 x_2 \theta_{7}/((\theta_{8} x_{6} + 1) M_1) - \theta_{7} x_{4}/M_1 - \theta_{11} x_8 x_{4} / \theta_{26} ,\\ 
\dot{x}_5& = &\theta_{7}  x_{4}/M_1 - \theta_{11} x_8 x_5 / \theta_{26} +  3\theta_{7} x_{3}/((\theta_{8} x_{6} + 1) M_1) ,\\ 
\dot{x}_6& = &-x_{6}  (\theta_{9} / \theta_{25})  (x_5 + x_{3}) ,\\ 
\dot{x}_7& = &\theta_{13}  x_8 - x_7 (\theta_{12} / \theta_{25})  x_{2345} ,\\ 
\dot{x}_8& = &x_7 (\theta_{12} / \theta_{25})  x_{2345} - \theta_{13}  x_8 ,\\ 
\dot{x}_9& = &k_6\theta_{23} x_{11}/k_7 - x_9 (\theta_{22} / \theta_{25})x_{2345}/M_1 - x_9 \theta_{21} (x_5 + x_{3})^2/((x_{18}  \theta_{3} / \theta_{1} + 1) M_1 \theta_{25}^2) ,\\ 
\dot{x}_{10}& = &x_9 \theta_{22} x_{2345} M_1/ \theta_{25} - \theta_{24} x_{10} + x_9 \theta_{21} (x_5 + x_{3})^2/ (\theta_{25}^2(x_{18}\theta_{3} / \theta_{1} + 1) M_1) ,\\ 
\dot{x}_{11}& = &k_7\theta_{24} x_{10}/k_6 - \theta_{23}  x_{11} ,\\ 
\dot{x}_{12}& = &-x_{12} \theta_{4} - \theta_{5}  x_{11}  (k_1 - 1)/ \theta_{27} ,\\ 
\dot{x}_{13}& = &x_{12}  \theta_{4} - x_{13}  \theta_{4} ,\\ 
\dot{x}_{14}& = &x_{13}  \theta_{4} - x_{14}  \theta_{4} ,\\ 
\dot{x}_{15}& = &x_{14}  \theta_{4} - x_{15}  \theta_{4} ,\\ 
\dot{x}_{16}& = &x_{15}  \theta_{4} - x_{16}  \theta_{4} ,\\ 
\dot{x}_{17}& = &x_{16}  \theta_{4}  k_6/k_7 - x_{17}  \theta_{5} ,\\ 
\dot{x}_{18}& = &x_{17}  \theta_{1}  \theta_{6} - x_{18}  \theta_{6} + k_2 \theta_{6} \theta_{2} \theta_{1} ,\\
\end{array}
\end{equation*}

\begin{equation*}\label{eqjak2}
\begin{array}{lll}
\dot{x}_{19}& = &- x_{19} \theta_{18} - \theta_{19}  x_{11}  (k_1 - 1)  / \theta_{27} ,\\ 
\dot{x}_{20}& = &x_{19}  \theta_{18} - x_{20}  \theta_{18} ,\\ 
\dot{x}_{21}& = &x_{20}  \theta_{18} - x_{21}  \theta_{18} ,\\ 
\dot{x}_{22}& = &x_{21}  \theta_{18} - x_{22}  \theta_{18} ,\\ 
\dot{x}_{23}& = &x_{22}  \theta_{18} - x_{23}  \theta_{18} ,\\ 
\dot{x}_{24}& = &k_6 x_{23} \theta_{18}/k_7 - x_{24}  \theta_{19} ,\\ 
\dot{x}_{25}& = &x_{24}  \theta_{15}  \theta_{20} - x_{25}  \theta_{20} + k_3 \theta_{20} \theta_{16} \theta_{15}
\end{array}
\end{equation*}
where the auxiliary variables $x_{2345} =x_2+x_3+x_4+x_5$ and $M_1 = x_{25}\theta_{17}/\theta_{15}+1$ have been used. 

The 25 states $x_1, x_2, \dots, x_{25}$ are, respectively, the following species: EpoRJAK2, EpoRpJAK2, p1EpoRpJAK2, p2EpoRpJAK2, p12EpoRpJAK2, EpoRJAK2\_CIS, SHP1, SHP1Act, STAT5, pSTAT5, npSTAT5, CISnRNA1, CISnRNA2, CISnRNA3, CISnRNA4, CISnRNA5, CISRNA, CIS, SOCS3nRNA1, SOCS3nRNA2, SOCS3nRNA3, SOCS3nRNA4, SOCS3nRNA5, SOCS3RNA, and SOCS3. 

The 27 unknown parameters, $\theta_i$, were written in the original publication \cite{bachmann2011division} as: CISEqc,
CISEqcOE,
CISInh,
CISRNADelay,
CISRNATurn,
CISTurn,
EpoRActJAK2,
EpoRCISInh,
EpoRCISRemove,
JAK2ActEpo,
JAK2EpoRDeaSHP1,
SHP1ActEpoR,
SHP1Dea,
SHP1ProOE,
SOCS3Eqc,
SOCS3EqcOE,
SOCS3Inh,
SOCS3RNADelay,
SOCS3RNATurn,
SOCS3Turn,
STAT5ActEpoR,
STAT5ActJAK2,
STAT5Exp,
STAT5Imp,
init\_EpoRJAK2,
init\_SHP1,
and
init\_STAT5.

The model has seven known constants ($k_1$--$k_7$), 5 of which correspond to experimental conditions that can be considered as constant inputs ($k_1$--$k_5$), including the external signal ($k_5 \equiv$ Epo).

The output equations are:
\begin{equation*}\label{outjak}
\begin{array}{lll}
y_{1} & = & 2(x_2 + x_3 + x_4 + x_5)\theta_{25},\\
y_{2} & = & 16(x_3 + x_4 + x_5)\theta_{25},\\
y_{3} & = & x_{18}\theta_1,\\
y_{4} & = & x_{25}/\theta_{14},\\
y_{5} & = & (x_9+x_{10})/\theta_{27},\\
y_{6} & = & x_{10}\theta_{27},\\
y_{7} & = & x_9,\\
y_{8} & = & x_7 + x_8,\\
y_{9} & = & x_{18},\\
y_{10} & = & x_{25},\\
y_{11} & = & 100 x_{10}/(x_{10}+x_9),\\
y_{12} & = & x_{24},\\
y_{13} & = & x_{17},\\
y_{14} & = & (x_7 + x_8) (1 + (k_4 \theta_{27})) / \theta_{26},\\
\end{array}
\end{equation*}


\end{document}